\documentclass[Journal,letterpaper]{ascelike-new}
\usepackage[utf8]{inputenc}
\usepackage[T1]{fontenc}
\usepackage{lmodern}
\usepackage{graphicx}
\usepackage[figurename=Fig.,labelfont=bf,labelsep=period]{caption}
\usepackage{subcaption}
\usepackage{amsmath}
\usepackage{comment}
\usepackage{newtxtext,newtxmath}
\usepackage[colorlinks=true,citecolor=red,linkcolor=black]{hyperref}
\NameTag{Guo, \today}
\providecommand{\keywords}[1]
{
  \small	
  \textbf{\textit{Keywords---}} #1
}

\begin{document}

\title{ORCLSim: A System Architecture for Studying Bicyclist and Pedestrian Physiological Behavior Through Immersive Virtual Environments}

\author[1]{Xiang Guo}
\author[1]{Austin Angulo}
\author[1]{Erin Robartes}
\author[1]{T. Donna Chen}
\author[1,*]{Arsalan Heydarian}

\affil[1]{Department of Engineering Systems and Environment, University of Virginia, Charlottesville, VA, 22904. *Email: ah6rx@virginia.edu}

\maketitle

\begin{abstract}
Injuries and fatalities for vulnerable road users, especially bicyclists and pedestrians, are on the rise. To better inform design for vulnerable road users, we need to conduct more studies to evaluate how bicyclist and pedestrian behavior and physiological states change in different roadway designs and contextual settings. Previous research highlights the advantages of Immersive Virtual Environment (IVE) in conducting bicyclist and pedestrian studies. These environments do not put participants at risk of getting injured, are low-cost compared to on-road or naturalistic studies and allow researchers to fully control variables of interest. In this paper, we propose a framework ORCLSim, to support human sensing techniques within IVE to evaluate bicyclist and pedestrian physiological and behavioral changes in different contextual settings. To showcase this framework, we present two case studies where we collect and analyze pilot data from five participants' physiological and behavioral responses in an IVE setting, representing real-world roadway segments and traffic conditions. Results from these case studies indicate that physiological data is sensitive to road environment changes and real-time events, especially changes in heart rate and gaze behavior. Additionally, our preliminary data indicates participants may respond differently to various roadway settings (e.g., intersections with or without traffic signal). By analyzing these changes, we can identify how participants' stress levels and cognitive load is impacted by the simulated surrounding environment. The ORCLSim system architecture can be further utilized for future studies in users' behavioral and physiological responses in different virtual reality settings. 
\end{abstract}


\section{Introduction}
Over the past couple of decades, the evaluation of roadway safety and design has been automobile-centric. Many observational, survey-based, naturalistic, and experimental studies have been conducted to evaluate the impact of roadway design features on drivers' behaviors and safety, leaving out other roadway users such as bicyclists and pedestrians. Furthermore, recent studies have highlighted the growth in the number of injuries and fatalities for vulnerable road users \cite{national2020fatality}. National Highway Traffic Safety Administration (NHTSA) reported a 35\% increase in pedestrian fatalities nationwide in the past ten years, and deaths of bicyclists in the United States reached all-time highs in 2018 and 2019. These reports also indicate that although the overall number of vehicle crash fatalities are decreasing (which includes vehicle and non-vehicle related), the proportion of vulnerable road users (motorcyclists, pedestrians, pedalcyclists, and other non-occupants in the vehicle) fatalities has been increasing from 20\% in 1996 to 34\% in 2019 \cite{NHTSA2020}. These trends indicate that the design of our roadways need to be improved to be more inclusive for all users, especially for the vulnerable road users such as bicyclists and pedestrians \cite{rodriguez2021towards}.

Different factors, such as the speed limit, roadway design, and presence of large vehicles (e.g., trucks) have been shown to be associated with severe injury or fatality of bicyclists \cite{chen2016built}. Additionally, the presence of intersections, traffic volumes, noise level, and physical segregation between bicyclists and vehicles have shown to influence bicyclists' stress or comfort level \cite{teixeira2020does,rybarczyk2020physiological,cobb2021bicyclists}. Similarly, for pedestrians' safety, similar factors as bicyclists' are emphasized by researchers: pedestrian infrastructure, roadway design, traffic volumes, vehicle speed, and visibility of the road environment \cite{stoker2015pedestrian}. It is also found that bicycle paths, crossing surface material, street type, as well as presence of nearby parked vehicles are associated with the number of pedestrian–vehicle conflicts from a naturalistic observation study \cite{cloutier2017outta}. 

To better inform roadway design, extensive datasets similar to the automobile-focused studies of the past are needed for bicyclists and pedestrians. To develop robust bicyclist and pedestrians-focused datasets, studies with both high ecological and internal validity are needed. Ecological validity refers to the extent an experimental environment matches with the real world, increasing the chances that the effects identified in an experimental environment generalize to real world settings. Internal validity refers to the extent in which a cause-effect relationship is warranted in a study. Subjective, naturalistic, and experimental datasets can be utilized to tackle these issues. Subjective studies, such as surveys, provide measures of users and their perceptions of their environment, but lack ecological and internal validity \cite{wynne2019systematic}. On the other hand, naturalistic studies can provide information about realistic changes within the environment and bicyclist and pedestrians' behavior with high ecological validity, but these studies with lower internal validity are resource- and time-extensive, and have potential risks of injuries and fatalities for participants. For example, a study in real traffic examining glance behavior of teenage cyclists while listening to music is terminated when the results indicated that a substantial percentage of participants cycling with music decreased their visual performance \cite{stelling2018study}. Furthermore, naturalistic studies are influenced by many environmental factors that restrict the ability to fully isolate and understand the impact of independent variables, thus offering low internal validity, especially for physiological and behavioral factors \cite{fitch2020psychological,teixeira2020does}. Thus far, the majority of bicyclist and pedestrian studies rely on subjective and naturalistic data derived from real-world settings to assess participants' behavior and comfort in different traffic environments \cite{van2019visual,shaaban2018analysis}. 

Experimental studies provide an opportunity to evaluate the impact of safety-related conditions, infrastructure, and technology on bicyclists and pedestrians as they can offer the ecological validity lacking in subjective studies and allow the researchers to control for external variables, unlike naturalistic studies, for greater internal validity. Experimental studies conducted with virtual simulators can minimize the hypothetical bias of subjective surveys while offering a controlled, low-risk, and immersive environment that real-world experiments cannot guarantee. The benefit of Immersive Virtual Environment (IVE) is achieving high internal and ecological validity while also being cost effective, and offering complete experimental control to replicate trials \cite{heydarian2015immersive,heydarian2017use}. Early IVE lacked realism, which was primarily due to a lack of technological capability. Fortunately, IVE software and hardware platforms have significantly improved over the last few years with the release of high end, commercially available head mounted displays (HMD), and graphics cards capable of rendering highly detailed environments. Meanwhile real-life components can also be integrated into IVE studies. For instance, we can now integrate a physical bicycle along with haptic feedback with an IVE to create a more immersive setting for participants \cite{guo_robartes_angulo_chen_heydarian_2021}. Furthermore, as the level of immersion increases, we can expand these experiments to capture participants' physiological and psychological factors, which is a field of data that has historically been overlooked. Such data provides insights into how participants behaviors and perceptions may change in contextual settings in different research fields \cite{kim2019saliency,lee2020wearable,adami2021effectiveness,noghabaei2021feasibility,sharif2021effect}. For example, pedestrians' distinct physiological responses (gait patterns, heart rate, and electrodermal activity) to negative environmental stimuli are reported from naturalistic ambulatory settings in a building \cite{kim2020environmental}. With the increase of realism in IVE simulators and the development of low-cost ubiquitous sensors, IVE simulators have become promising tools for conducting highly realistic and immersive experimental studies \cite{ergan2019quantifying}. In traffic safety studies, driving simulators have been widely applied to study drivers’ behaviors, awareness \cite{soares2020driving}, and psychophysiological states with multimodal data collection systems, such as eye trackers, electroencephalogram (EEG) and electrocardiogram (ECG) \cite{haufe2011eeg,akbar2017three,guo2019will,baee2021medirl}. Some of the driver-related studies are conducted in IVE \cite{eudave2017physiological}. Meanwhile, for bicyclists and pedestrians, only a few studies have applied physiological responses in IVE simulators. For example, bicyclists' galvanic skin response had less peaks with a bike lane than no bike lane condition \cite{cobb2021bicyclists}. In another cycling virtual reality study, electroencephalography (EEG) data shows its potential in a hybrid model framework as an indicator of perceived risk of bicyclists \cite{bogacz2021modelling}.
For pedestrians, it is notable that older pedestrians spent more time focusing on the central area of the scene and even less so in the last five seconds before making the crossing decision in an IVE study \cite{tapiro2016older}. 

In this paper, we propose a IVE-based framework (ORCLSim) for supporting pedestrian and bicyclist research. The proposed framework integrates realistic visualizations from the real world in IVE along with a physical bicycle and a suite of passive sensing technologies, which enable the collection of physiological and behavioral responses of users. The goals of this paper are to:
1. identify research methods, trends, and gaps in knowledge related to bicyclist and pedestrian research in IVE; 2. present a novel framework for evaluating bicyclist and pedestrian behavioral changes through integrating human physiological sensing within IVE; 3. present a set of case studies to highlight how the proposed framework could be implemented to collect and analyze bicyclist and pedestrians' behavioral and physiological changes in different roadway conditions and designs. 

\section{Background and Gaps in IVE Simulator Knowledge}
This section will provide background information regarding different types of studies on bicyclists and pedestrians and how physiological measurement are integrated in related studies, especially for IVE studies.

\subsection{Surveys and Observational Studies}
Surveys have been widely used as methods of studying bicyclists and pedestrians, particularly when faced with a lack of real-world data. Surveys, when composed carefully, can reliably and efficiently assess large populations of people and have been used to study a wide variety of topics including: perceived safety/comfort \cite{parkin2007models,chaurand2013cyclists,abadi2018bicyclist}, route choice \cite{sener2009analysis} and crash history \cite{robartes2018crash,poulos2012exposure,yang2019underreporting}. However, stated preference surveys have limitations, such as being subject to hypothetical bias where responses to hypothetical situations are not the same as they would be in real-world situations \cite{fitch2018relationship}. 

Observational studies eliminate the risk of hypothetical bias from stated preference surveys \cite{thompson2013impact}. However, the data collected relies on road users in real-world conditions. In recent years, with the increasing number of cameras, more video streams are available for observational studies. However, these studies can only evaluate participants' behaviors in existing environments, where we have very limited number of options to consider for roadway improvements. To have full control over design considerations, we need to evaluate how bicyclists and pedestrians respond to different design of roadways during the planning or design phase of projects. Simulations and immersive virtual environments offer an approach which minimizes the limitations of stated preference surveys, and allows for a controlled, safe environment that real world observational studies cannot provide.

\subsection{IVE Simulation Technology and Framework}
 Over the past decade, driving simulators, virtual reality (VR) technologies, and human sensing technologies have provided new insights on human behavior in different contextual settings, assisting in evaluating different design alternatives for roadways \cite{subramanian2021pedestrians,zou2021road}, buildings \cite{francisco2018occupant}, hospitals \cite{chias20193d}, and other civil infrastructure systems \cite{zou2017emotional,noghabaei2020hazard,awada2021integrated}. Simulation methods utilizing IVE offer a low-cost, low-risk approach to studying the users' safety, perception, and behavior. Traditionally, real-world observation methods have been used to understand bicyclist and pedestrian behavior. These methods are often expensive, time consuming, and unrealistic for studying naturalistic behaviors as they often require some level of unrealistic environmental control for the safety of test subjects. The improvements in IVE over the recent years has provided researchers, designers, and engineers a way to evaluate alternative infrastructure design while providing high degrees of immersion. Novel, commercially available VR headsets offer a high degree of realism and immersion. Furthermore, environmental factors that may influence bicyclist and pedestrian behavior are highly controllable within IVE, allowing for replicable experimental trials. The last two decades have seen research utilizing IVE and VR simulations focusing on how countermeasures influence safety-related elements such as walking speed, gap acceptance, analysis of risky behavior, stated preference data, visual or auditory warning effectiveness, speeds, steering, and resistance have all been the commonly discussed in previous research \cite{neider2010pedestrians,stevens2013preadolescent,deb2017efficacy,o2017validation,brown2017investigation,sun2018design}. 

Arguably the biggest gap in IVE research for bicyclists and pedestrians is the lack of standard methods to cross-compare different studies. For instance, it is difficult to draw conclusions relating to technology effectiveness between a simulator using 2D screens and another using a 3D HMD as validation studies are very limited and not consistent between different mediums \cite{schneider2020virtually}. This was shown by Maillot et al. (2017) in which they evaluated participants' crossing behavior across three mediums: 2D screens, 3D HMD, and 3D Cave Automatic Virtual Environment (CAVE); their analysis showed there exists a significant difference in participant gap acceptance between 2D screens and CAVE. However, there was no significant difference between CAVE and HMDs \cite{maillot2017training,mallaro2017comparison}. This indicates that not only there are limited IVE studies for understanding bicyclist and pedestrian behaviors, there is also a very limited number of studies that their results and findings can be properly compared. Some correlation is recognized between a multiscreen setup and the use of a cell phone mounted in a cardboard viewer as a simulated HMD setup \cite{schwebel2017using}. Other comparisons have been taken into consideration of the fidelity include participant movement, visual scenes, and the sound technology \cite{montuwy2016adapting,jiang2017influence,bernhard2008bimodal,tzanavari2014user}. Overall, a lack of generalized framework to develop IVE simulators and technological inconsistency in data collection between studies are the biggest factors for this gap. Other gaps worth noting include a lack of model complexity; for example, more work needs to be put into the IVEs incorporating traffic flow theory   \cite{farooq2018virtual,deb2018investigating,bhagavathula2018reality}. Additionally, a lack of complexity with respect to what the bicyclists and pedestrians can do within an IVE also needs to be addressed, including limitations in walking speed, interaction with vehicles and infrastructure, and modeling streetscapes within the boundaries of indoor laboratory space \cite{deb2017efficacy,deb2018investigating,farooq2018virtual,bhagavathula2018reality}. 

\subsection{Integration of Human Sensing within IVE}
Apart from subjective studies, there are limited datasets including human physiological and psychological sensing (e.g. eye tracking, body tracking and heart rate) for bicyclists and pedestrians. It is crucial to assess participants' patterns of perception and reaction in certain contextual settings. Many traditional on-road studies have used accident statistics and road infrastructure data (e.g., road-side cameras) to evaluate the safety-related concerns of bicyclists and pedestrians. To further study their perception and cognitive states, human sensing devices (e.g. physiology devices) have shown to provide promising insights \cite{ridel2018literature,trefzger2018visual,ayres2015bicyclist}. There are practical concerns about the data collection of human sensing on real roads. First, the safety, ethical and cost considerations prohibit large-scale on-road experiments \cite{stelling2018study}. Second, the implementation of traditional human sensing devices (such as body trackers) are intrusive, which may affect the behavioral and perception ability of the participant, as well as the data quality on real roads (especially in high speed scenarios) \cite{van2012comparing}. 

In light of these shortcomings, most IVE are able to handle the first limitation, as virtual environments provide a low-risk and cost-effective alternative to the real settings. While the second shortcoming (monitoring perception and cognition) requires integration of human sensing systems and ubiquitous computing into the IVE. The majority of existing IVE research in bicyclist and pedestrian studies have not utilized ubiquitous computing and human sensing techniques to monitor participants' behaviors and physiological states. 

Eye tracking behaviors, such as fixations distribution and pupillometry, are usually found to be related to the process of cognitive resource allocation. Eye-tracking behavior is usually measured by optical eye-trackers. Eye-tracking has been widely used in studying users' visual perception and attention in different contexts. For example, research has shown that experienced and inexperienced bicyclists have a different perceived gaze at infrastructure treatments around intersections \cite{rupi2019visual}. The latest virtual reality headsets, such as the HTC VIVE Pro Eye, have integrated eye tracking features, allowing for IVE researchers to incorporate eye tracking analysis within their studies.

Body position has an influence on leg kinematics and muscle recruitment for bicyclists \cite{chapman2008patterns}. Some professional sensors can be used to build 3D body tracking by implementing multiple on-body receivers to study pedestrians’ dynamics of indoor activity \cite{correa2016indoor}. Recent development in computer vision has greatly reduced the cost of obtaining body movement data. For example, OpenPose, an open source real-time multi-person system, is able to jointly detect human body, hand, facial, and feet key points on single 2D images \cite{cao2017realtime}. 

Electrocardiography (ECG) is a well-established method to record the electrical activity of the heart. A participant’s heart rate (HR) and heart rate variability (HRV) can be measured using an ECG signal. HR is a commonly measured index of physiological arousal in response to changes in working demands, especially for workload \cite{jorna1992spectral}. Relative to HR, HRV decreases with increasing task demands \cite{hoover2012real}. To collect the HR/HRV data, apart from the intrusive sensors usually utilized in lab tests, many wearable devices, such as smartwatches and smart bands, can provide reliable measurement for HR and HRV \cite{exler2016wearable,tavakoli2021driver}. These devices enable longitudinal data collection which can help in building personalized models for users.

To summarize existing literature in this area, we have categorized past IVE bicyclist and pedestrian simulator studies with their IVE settings and data collection methods. Table \ref{bicycle literature table} and \ref{pedestrian literature table} have been developed to better illustrate how the trends in technology, immersion, collected data, and analysis of bicyclist and pedestrian research have changed over the last two decades. Note that for studies from the same research group, only the latest work is included. 

The work detailed in this paper addresses the identified gaps (1) by providing a framework in developing highly immersive IVE to evaluate the road designs, and (2) through the use of commercially available technology to collect multimodal human sensing data to study bicyclist and pedestrian physiological and behavioral responses in different road environments.

\begin{table*}[]
\caption{IVE bicyclist simulator literature table. \textbf{-}: not included or not specified in the paper; \textbf{\checkmark}: included in the paper. \textbf{Visual Technology}: Subject viewed a \textit{single screen/multiple screen or CAVE/head mounted display(HMD)} as visual source; \textbf{Agency of Movement}: \textit{Stationary} - subject remained motionless or interacted via controller; \textit{Dummy} - subject was on stationary bike but movements were not translated into VR; \textit{Real-time} - subject movements were translated in VR. \textbf{Sound}: whether sound feedback was used. \textbf{Haptic}: Interaction with the environment through, vibration, resistance, etc. \textbf{Kinematic}: speed, steering and direction data. \textbf{Movement}: body or head movements.}
\label{bicycle literature table}
\resizebox{\textwidth}{!}{
\begin{tabular}{|c|c|c|c|c|c|c|c|c|c|c|c|c|}
\hline
\multicolumn{2}{|c|}{\textbf{Report Information}}
 & \multicolumn{5}{|c|}{\textbf{Level of Immersion}} 
 & \multicolumn{6}{|c|}{\textbf{Data Reported}} 
 \\
 \hline
 \textbf{Author \& Year} & \textbf{Laboratory or Affiliation} & \textbf{Simulator Environment Setting} & \textbf{Visual Technology}
 & \textbf{Agency of Movement} & \textbf{Sound} & \textbf{Haptic} & \rotatebox[origin=c]{90}{\textbf{Participant Number}} & \rotatebox[origin=c]{90}{\textbf{Kinematic}}  & \rotatebox[origin=c]{90}{\textbf{Movement}} & \rotatebox[origin=c]{90}{\textbf{Eye Tracking}} & \rotatebox[origin=c]{90}{\textbf{Physiological}}  & \rotatebox[origin=c]{90}{\textbf{Stated Preference}} \\
 \hline
 \hline
 \cite{van1998navigating} & Max-Planck-lnstitute for Biological Cybernetics & Real-world (Tübingen, Germany) &  Single Screen + HMD & Real-time & - & \checkmark & - & \checkmark & - &
 - & - & - \\
\hline
 \cite{kwon2001kaist} & KAIST Bicycle Simulator & Real-world (KAIST Campus, Korea) & Single Screen + HMD & Real-time & - & \checkmark & - & \checkmark & \checkmark &
 - & - & - \\
\hline
 \cite{nikolas2016risky} & Hank Virtual Environments Lab & Simulation & CAVE & Real-time & \checkmark & \checkmark & 63 & \checkmark & - &
 - & - & \checkmark \\
\hline
 \cite{o2017validation} & Monash University Accident Research Centre& Real-world (Monash University)  & HMD & Dummy & - & - & 30 & \checkmark & - &
 - & - & \checkmark \\ 
\hline
 \cite{xu2017exploring} & Intelligent Human Machine Systems Lab  & Simulation & HMD & Stationary & - & - & 30 & \checkmark & - &
 - & - & \checkmark \\  
\hline
 \cite{kwigizile2017real} & Western  Michigan  University  & Real-world (Western Michigan University Campus) & HMD & Real-time & - & \checkmark & 36 & \checkmark & - &
 - & \checkmark & \checkmark \\  
 \hline
 \cite{lee2017description} & Delft University of Technology & - & HMD & Real-time & - & - & - & \checkmark & - &
 - & - & - \\  
 \hline
 \cite{stroh2017design} & University of Iowa & -  & CAVE & Real-time & - & - & - & \checkmark & - &
 - & - & - \\   
 \hline
 \cite{keler2018bicycle} & Technical University Munich  & Real-world (Munich, Germany) & Single Screen + HMD & Real-time & - & - & - & \checkmark & - &- & - & - \\  
   \hline
  \cite{sun2018design} & ZouSim, University of Missouri  & Real-world (Columbia, Missouri) & CAVE + HMD & Real-time & - & \checkmark & - & \checkmark & - &
 - & - & - \\  
 \hline
 \cite{nazemi2018speed} & Future Cities Laboratory & - & HMD & Dummy & - & - & - & - &
 - & - & - &- \\
\hline
\cite{abadi2019factors} & Oregon State University & - & Single Screen & Real-time & \checkmark & - & 48 & - & - &
 - & - & \checkmark \\
\hline
\cite{shoman2020modeling} & IFSTTAR & - & CAVE & Real-time & - & \checkmark & 10 & \checkmark & - &
 - & - & \checkmark \\
\hline
Current study & ORCL, University of Virginia & Real-world (Charlottesville, Virginia) & HMD & Real-time & \checkmark & \checkmark & - & \checkmark & \checkmark &
 \checkmark & \checkmark & \checkmark \\
\hline
 
\end{tabular}
}
\end{table*}

\begin{table*}[]
\caption{IVE pedestrian simulator literature table. \textbf{-}: not included or not specified in the paper; \textbf{\checkmark}: included in the paper. \textbf{Visual Technology}: Subject viewed a \textit{single screen/multiple screen or CAVE/head mounted display(HMD)} as visual source; \textbf{Agency of Movement}: \textit{Stationary} - subject remained motionless or interacted via controller; \textit{Dummy} - subject walked on treadmill or stepped off platform but actions weren't translated in VR, movement was only proxy; \textit{Real-time} - subject movements were translated in VR. \textbf{Sound}: whether sound feedback was used. \textbf{Haptic}: Interaction with the environment through, vibration, resistance, etc. \textbf{Kinematic}: speed, steering and direction data. \textbf{Movement}: body or head movements.}
\label{pedestrian literature table}
\resizebox{\textwidth}{!}{
\begin{tabular}{|c|c|c|c|c|c|c|c|c|c|c|c|c|}
\hline
\multicolumn{2}{|c|}{\textbf{Report Information}}
 & \multicolumn{5}{|c|}{\textbf{Level of Immersion}} 
 & \multicolumn{6}{|c|}{\textbf{Data Reported}} 
  \\
  \hline
 \textbf{Author \& Year} & \textbf{Laboratory or Affiliation} & \textbf{Simulator Environment Setting} & \textbf{Visual Technology}
 & \textbf{Agency of Movement} & \textbf{Sound} & \textbf{Haptic} & \rotatebox[origin=c]{90}{\textbf{Participant Number}} & \rotatebox[origin=c]{90}{\textbf{Kinematic}}  & \rotatebox[origin=c]{90}{\textbf{Movement}} & \rotatebox[origin=c]{90}{\textbf{Eye Tracking}} & \rotatebox[origin=c]{90}{\textbf{Physiological}}  & \rotatebox[origin=c]{90}{\textbf{Stated Preference}} \\
 \hline
 \hline
 \cite{simpson2003investigation} & University of Canterbury & Simulation & HMD & Real-time & - & - & 24 &- & \checkmark & - & - & - \\
 
 \hline
  \cite{banton2005perception} & University of Virginia & - & HMD & Dummy & - & - & 57 & \checkmark & \checkmark & - & - & \checkmark \\
 
  \hline
  \cite{tzanavari2015effectiveness} & Immersive and Creative Technologies Lab  & - & CAVE & Real-time & \checkmark & - & 6 & - & \checkmark & - & - & - \\

 \hline
\cite{schwebel2017experiential} & UAB Youth Safety Lab & Real world (University of Alabama at Birmingham) & Dummy & Stationary & - & - & 219 & - & \checkmark & - & - &\checkmark \\

\hline
\cite{mallaro2017comparison} & University of Iowa & Simulation & CAVE + HMD & Real-time & \checkmark & - & 32 & - & \checkmark & - & - & - \\

  \hline
\cite{maillot2017training} & Ifsttar & Simulation & CAVE & Stationary+ Real-time & - & - & 20+40 & - & \checkmark & - & - & - \\

  \hline
\cite{iryo2018applicability} & Nagoya University & Simulation & HMD & Real-time & \checkmark & - & 32 & \checkmark & - & - & - &\checkmark \\

  \hline
\cite{iryo2018applicability} & Nagoya University & Simulation & HMD & Real-time & \checkmark & - & 32 & \checkmark & - & - & - &\checkmark \\

  \hline
\cite{farooq2018virtual} & Laboratory of Innovations in Transportation & Simulation & HMD & Real-time & \checkmark & - & 42 & - & - & - & - &\checkmark \\

  \hline
\cite{deb2018investigating} & Mississippi State University & Simulation & HMD & Real-time & \checkmark & - & 30 & \checkmark & \checkmark & - & - &\checkmark \\

  \hline
\cite{bhagavathula2018reality} & Virginia Tech Transportation Institute & Real world(Virginia Smart Road) & HMD & Dummy & - & - & 16 & \checkmark & - & - & - &\checkmark \\

  \hline
\cite{cavallo2019street} & Ifsttar  & - & CAVE & Real-time & \checkmark & - & 79 & - & \checkmark & - & - & - \\

\hline
Current study & ORCL, University of Virginia & Real-world (Charlottesville, Virginia) & HMD & Real-time & \checkmark & \checkmark & - & \checkmark & \checkmark &
 \checkmark & \checkmark & \checkmark \\
\hline
\end{tabular}
}
\end{table*}

\section{Methodology}

\begin{figure}
\centering
\includegraphics[width=\textwidth]{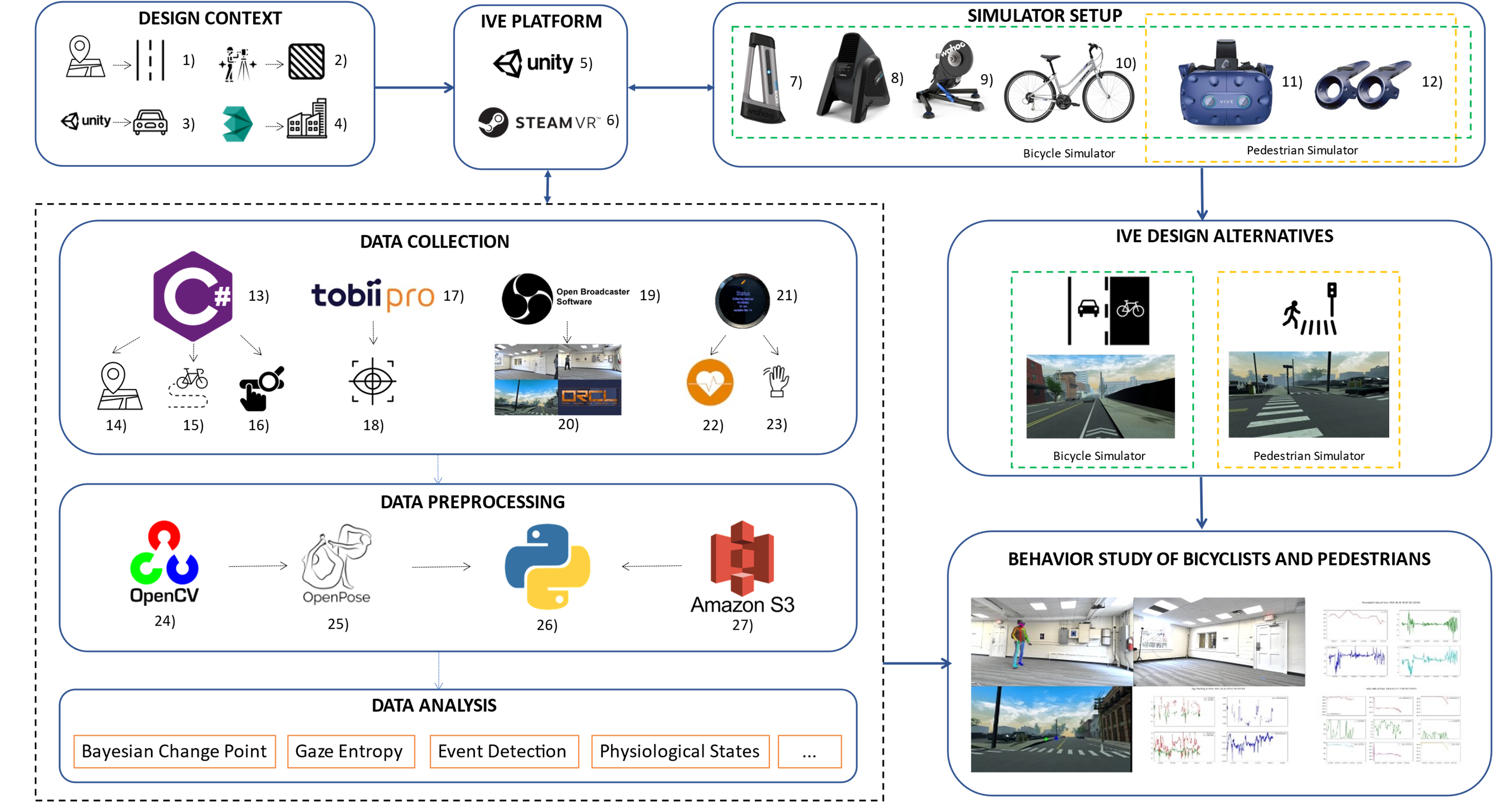}
\caption{System architecture of ORCLSim framework. Design context: 1) Road geometry information from Google map; 2) Road texture from real world measurement; 3) Vehicle modeling and traffic simulation in Unity; 4) Buildings modeling from 3DMax. IVE Platform:  5) Unity: 3D gaming engine; 6) SteamVR: integrating hardware with Unity. Simulator Setup: 7) Wahoo Kickr Climer: physical grade changes; 8)  Wahoo Kickr headwind: headwind simulation by speed; 9) Wahoo Kickr Smart Trianer + ANT+: biking dynamics simulation; 10) Trek Verve physical bike; 11) HTC VIVE Pro Eye: VR headset with eye tracking; 12) Controllers: steering and braking of the bike and pedestrian’s interactions with the environment. Data Collection: 13) C\# scripts in Unity to record: 14) Position, 15) Cycling performance and 16) Pedestrian’s interactions (touch, click or press); 17) TobiiPro Unity API collects: 18) Eye tracking data; 19) OBS studio: records room videos and VR videos simultaneously as shown in 20); 21) Android smartwatch collects: 22) Heart rate and 23) hand acceleration data. Data Preprocessing:  24) Opencv: video and image processing; 25) Openpose: pose data extraction from videos; 26) Python: data cleaning, management and analysis;  27) Amazon S3: smartwatch data on the cloud}
\label{fig:system_archetecture}
\end{figure}

To address the existing knowledge gaps identified in the previous section, we introduce a new IVE-based framework - ORCLSim where we can evaluate participants' behavioral and physiological responses in different simulated environments. This section provides details on the devices and processing techniques utilized in the proposed framework. In order to collect the multimodal data desired, multiple components are required to work in synchronicity within the IVE. The ORCLSim system architecture is shown in Fig.\ref{fig:system_archetecture}, detailing all of the technology, software, communications network, and associated data flow. The details of the system framework will be discussed in this section.

\subsection{Environment and Design Context}

\begin{figure}
\centering
\includegraphics[width=\textwidth]{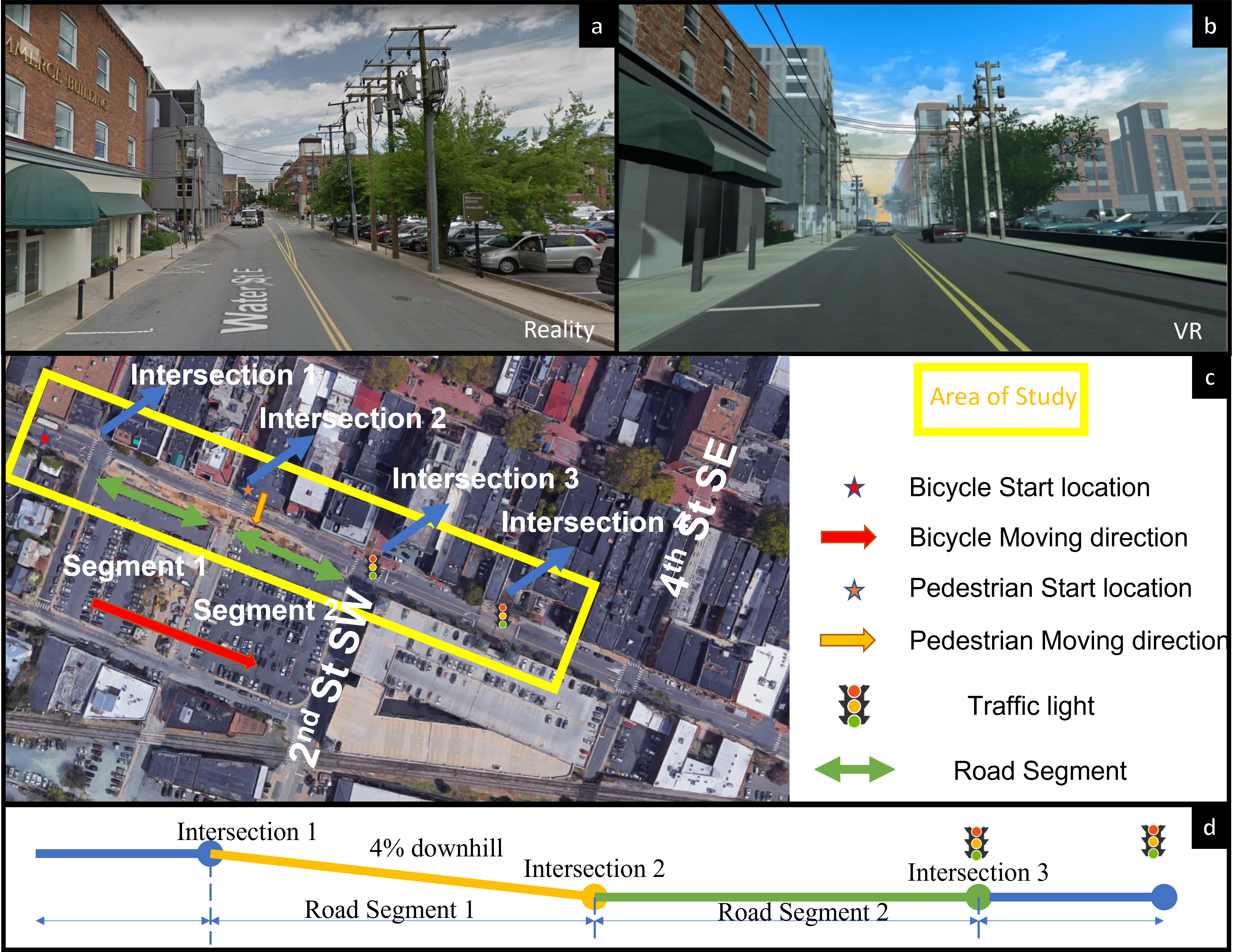}
\caption{Comparison between (a) the real environment street map and (b) the IVE, (c) the map for the location of the real world environment, (d) profile of road geometry}
\label{fig:location}
\end{figure}

The IVE is developed based on a real-world location: the Water Street corridor in Charlottesville, Virginia. Water Street is well-trafficked by bicyclists, and has been identified by Virginia Department of Transportation as a high risk site for pedestrians, and is being considered for redesign by the city of Charlottesville as shown in Fig.\ref{fig:location}. The section of the corridor chosen for this experiment consists of four city blocks, with a 4\% eastbound downhill in one of the road segment(road segment 1 in Fig.\ref{fig:location}d), shared lane markings for bicycles in the east and westbound directions, a traffic signal at the intersection of East Water Street and 2nd Street SE, and a parking lane in the westbound direction. The IVE was developed on a one-to-one scale of the Water Street corridor based on technical drawings provided by the City of Charlottesville and in-field measurements (Fig.\ref{fig:system_archetecture}-1). The textures - graphical images/skins laid atop 3D models to represent surface detail - used within the IVE were custom made from high resolution images taken on-site of the real-world surfaces appearing in the IVE so that all colors and surface details within the IVE represented the exact same of those in the real-world environment (Fig.\ref{fig:system_archetecture}-2). Fig.\ref{fig:location} (a) and (b) present the comparison between the real environment and the IVE created in Unity. 

Real world observations were conducted by installing four MioVision Scout cameras at the intersections of 2nd St SW, 1st St S, and 2nd St SE along Water St. The cameras captured video footage for two time periods: Tuesday 12 am to Thursday 11:59 pm on August 27-29 and September 3-5, 2019. Midweek days were chosen to avoid any abnormal fluctuations in traffic typically seen on Mondays and Fridays. Vehicle traffic within the IVE is generated based on the observed vehicle traffic during these time periods. A cumulative distribution function (CDF) was developed for the observed headway gaps between vehicles in both directions of traffic. The resulting CDF was used to generate multiple, randomly weighted theoretical gap observations. The order of presentation of the gaps from the resulting theoretical distribution were randomized during each experiment to avoid any bias towards gap sizes or learning effects. Furthermore, four different car models were used within the IVE, and each car model was randomly chosen each time for vehicles generated within IVE to further limit subject bias towards certain vehicles or learning effects (Fig.\ref{fig:system_archetecture}-3). The buildings in the IVE are modelled individually in 3DMax and then imported into Unity (Fig.\ref{fig:system_archetecture}-4). With all these methods aforementioned, we aim to maximize the immersion of the IVE.

\subsection{IVE platforms}
The IVE used in this framework is developed in Unity 3D game engine 2018 and run through the SteamVR platform. High end computing equipment is chosen so that development and testing would not be limited by computational performance. Two high-performance, factory overclocked Nvidia 1080Ti graphics cards run through Scalable Link Interface, an Intel Core i9-7920X CPU, 64 GB of DDR4 RAM at clock speeds of 3600MHz, and M.2 Solid State Hard Drives are installed in the lab computer to assure that environment rendering and stability, data collection speed, and information exchanges would not bottleneck at any component within the system while running SteamVR and Unity. 

\subsection{Simulator setup}
\begin{figure}
\centering
\includegraphics[width=\textwidth]{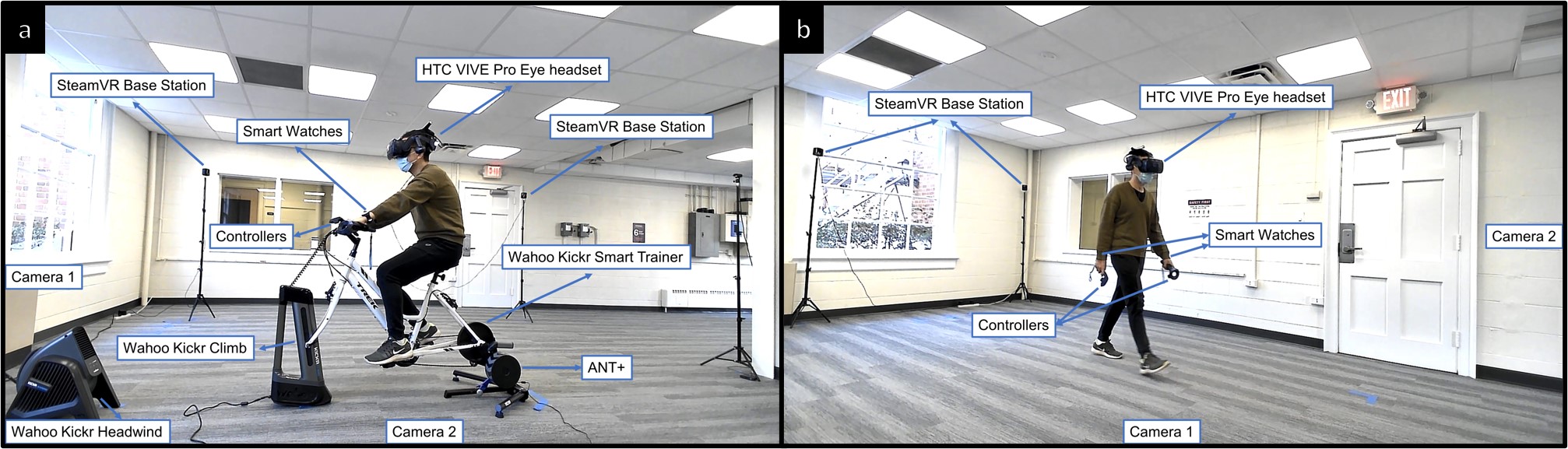}
\caption{Appearance of the simulators, (a) Bicycle simulator; (b) Pedestrian simulator}
\label{fig:Simulators}
\end{figure}

This section will discuss the hardware components chosen for both simulators. Fig.\ref{fig:Simulators} demonstrates the the appearances of both simulators. HTC Vive Pro VR headsets (Fig.\ref{fig:system_archetecture}-11) with their accompanying controllers (Fig.\ref{fig:system_archetecture}-12) are equipped in our simulators. The HTC Vive Pro is capable of running high resolutions (1440 x 1600 pixels per eye) and frame rates (90Hz), provides a wide field of view (110 degree), has movement tracing and gaze tracking capabilities, and is compatible with SteamVR. Controllers can be used differently in bicyclist and pedestrian studies. For bicyclists, the spatial location of the controllers allows the system to detect turning movements. The braking action can be recognized by the squeezing value of the trigger keys on the controllers. For pedestrian studies, controllers can help the user to interact with objects in the virtual environment, as an extension of their hands.  

\subsubsection{Bicycle Specific Equipment}
The following equipment has been specifically chosen for the bicyclist simulator:
\begin{itemize}
    \item Wahoo Kickr Smart Trainer (Fig.\ref{fig:system_archetecture}-9)  - Power measurement system of +/- 2\% for accurate, realistic resistance feedback. It has adaptive, real-time resistance based on road grade.
    \item Wahoo Kickr Climb (Fig.\ref{fig:system_archetecture}-7)  - Adaptive, real-time indoor bicycle grade simulator attached to the front fork of the bicycle that accurately raises or lowers the front end of the bicycle based on road grade. It is capable of simulating roadway grade in a range of -10\% to +20\%.
    \item Wahoo Kickr Headwind (Fig.\ref{fig:system_archetecture}-8)  - It provides adaptive, real-time variable speed vortex fan capable of reaching wind speeds experienced by bicyclists on the road, providing tactile feedback based on bicyclist speed.
    \item ANT+ (Fig.\ref{fig:system_archetecture}-9)  - Wireless protocol used for communications between the Wahoo training equipment and desktop computer, capable of sending controller information between the two devices regarding speed, resistance, grade, and wind speed.
\end{itemize}
The Wahoo equipment collects the data necessary for research studies and provides haptic feedback to the users. Critically, through the use of ANT+ and a Unity asset, the Wahoo equipment is compatible with the computer hardware and environment software used in the simulators. An average physical Trek Verve bike (Fig.\ref{fig:system_archetecture}-10) is adapted as the main body structure of the bicycle simulator.

\subsubsection{Pedestrian Specific Equipment}
The HTC Vive Pro Eye headsets have been equipped with the HTC Vive Pro Wireless Adapters, which support a 6 x 6 m space for accurate tracking and operating on a wireless communication. This HTC Wireless Adapter is chosen to meet the goals of the simulator as it eliminates the impact that wires hanging from the back of the headsets have on the users' ability to move around without getting tangled.

\subsection{Data collection}
In this section, we will introduce details about the collected data from different data sources, including the data type and the frequency of data collection. Specifically, we first discuss about the data streams exported from the Unity software, followed by the eye tracking data points, information extracted from the video recordings and smartwatches, as shown in Fig.\ref{fig:system_archetecture}. 

\subsubsection{Unity}
With the attached scripts written in C\# programming language to the Unity scenario (Fig.\ref{fig:system_archetecture}-13), we are able to extract the world position (in meters) and direction (unit vector) of each object in the virtual environment, including headset, controllers and other virtual objects such as vehicles. This data can decipher where the users are in the environment and their relative positions to other objects. The scripts also collect any input from the controllers. For example, the pulled trigger values (0 to 1) is the brake for the bike simulator. The frequency of Unity is generally around 30Hz. Additionally the system timestamp is attached to the final Unity output data for time synchronization. 

\subsubsection{Eye tracking}
The eye tracking features of HTC VIVE Eye Pro in Unity is from the integrated Tobii Pro eye tracker. The raw data extracted from the headset can be processed to track and analyze eye movement, attention, and focus level of each participant. Details of the utilized eye tracking system, sample environment and the code to extract the different data streams have been shared online \cite{xiangwebsite}. The eye tracking data is collected through Tobii Pro Unity SDK (Fig.\ref{fig:system_archetecture}-17). It is integrated in Unity with C\# scripts, and the data collection starts simultaneously with the running of the Unity scenes. The system timestamp is attached in the final output data. The output of Tobii Pro raw data is the 3D gaze direction, gaze origin, and pupil diameter. Pre-processing techniques are required to relate the eye tracker's coordinate system to the headset's position in a virtual 3D world (such as Unity). The frequency of eye tracking data is 120Hz. 

\subsubsection{Video Recording}
The video recording system has three components: two video recordings from cameras capturing the body position of the participant and one screen recording of the participant’s point of view in IVE. These videos are recorded simultaneously in OBS studio  (Fig.\ref{fig:system_archetecture}-19) with the same frequency (30Hz), resolution (1080p, 1920 $\times$ 1080) and system timestamp. 

\subsubsection{Smartwatch}
Experiment participants wear two android smartwatches ( (Fig.\ref{fig:system_archetecture}-21, one for each wrist) that are equipped with the “SWEAR” app for collecting longitudinal data. The SWEAR app records heart rate (1 Hz), hand acceleration (10 Hz), audio amplitude (noise level, 1/60 Hz), and gyroscope (10 Hz) \cite{boukhechba2020swear}. Both watches are connected to a smartphone via Bluetooth, the smartphone and computer are on the same wifi network to make sure time is synchronized with the server before each experiment. All data from the smartwatches are stored on the local device and then are uploaded to the Amazon S3 cloud  (Fig.\ref{fig:system_archetecture}-27) for future data extraction and analysis.

\subsection{Data Pre-processing and Integration}
\begin{figure}
\centering
\includegraphics[width=\textwidth]{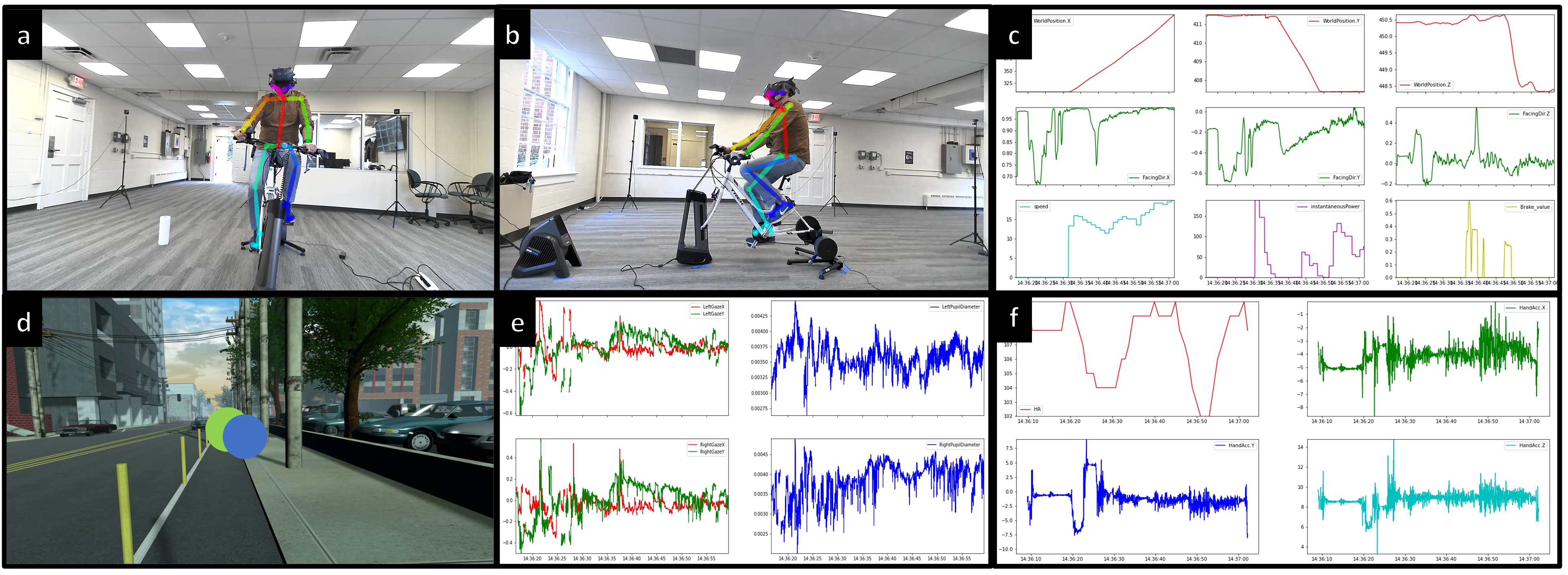}
\caption{Example of data visualization output. (a,b) Video data from two cameras with body position detection; (c) position and controllers input data in VR ; (d) Field of view VR video recording with gaze mapping, green/blue dots indicate left/right eye fixation; (e) Eye tracking data, includes gaze direction and pupil diameter; (f) Heart rate and hand acceleration data from smart watch}
\label{fig:data_visualization}
\end{figure}

All the data collection devices and platforms (except for the smartwatches) are connected to the local computer, allowing them synchronized with the computer's system time.  Fig.\ref{fig:data_visualization} shows the visualization of all the data collected in the simulator after time synchronization.

Information of each video source (frames per second, creation date, duration, height, and width) can be extracted from the singular video and be split into separate videos for each source (cameras 1 and 2) through the Opencv software (Fig.\ref{fig:system_archetecture}-24). Two external video cameras in the lab capture each participant’s body movement during the experiment. These recordings can be used to understand participants' movement and reactions. Furthermore, the body position data can be extracted from these videos using the OpenPose software (Fig.\ref{fig:system_archetecture}-25). 
Fig.\ref{fig:data_visualization}(a) and (b) show the body position detection of the video recordings from OpenPose. Used in conjunction with the movement tracking of the VR headset and controllers, this video footage can help determine how participants react to their environments during experimentation for better behavioral analysis and event detection.

Combining the raw gaze direction from the eye tracking data with the video information of the point-of-view videos, it is possible to transform the 3D gaze direction into 2D videos to visualize what the participants are looking at in the IVE. As shown in Fig.\ref{fig:data_visualization}(d), the green and blue dots represent the direction left and right eyes are looking at, respectively. 

Data from the smartwatches are stored locally in the device during the experiment and then uploaded to the Amazon S3 cloud storage. After the experiment, the data can be downloaded for further analysis. 

All the text data are transformed to .csv format for easier integration. All the physiological data are labeled corresponding to different road segments based on the time and position data from the unity output, where the road geometry information comes from.

\subsection{Data analysis}

In this section we discuss the change point detection algorithm applied on the HR as well as the gaze entropy, which is the basis of the event detection in our study. First, we discuss how the Bayesian change point (BCP) detection is applied on the HR data. Similarly, for the gaze data, we discuss how gaze entropy can be calculated and used to  identify the dispersion of gaze. 

\subsubsection{Bayesian Change Point Detection}
Bayesian Change Point (BCP) detection methods are applied to detect the abrupt changes in HR data. Change point analysis deals with time series data where certain characteristics undergo occasional changes. It is assumed that there is an underlying sequence of parameters partitioned into contiguous blocks of equal parameter values: the beginning of each block is a change point. Observations are then assumed to be independent in different blocks given the sequence of parameters \cite{barry1993bayesian}. A Bayesian approach to the change point problem can give uncertainty estimates not only for location, but also for the number of change points. 

Suppose we have a time series of HR data X, and we use $\rho = (U_1,...,U_n)$ to indicate a partition of the time series into non-overlapping HR regimes where $U_i=1$ means a change point happens at position $i$+1. To calculate the posterior distribution over partitions, we use the Markov Chain Monte Carlo (MCMC) method. We define a Markov Chain with the following transition rule: with probability $p_i$, a new change point at the location $i$ is introduced. In each step of the Markov Chain, at each position $i$, a value of $U_i$ is drawn from the conditional distribution of $U_i$ given
the data X and the current partition $\rho$. let $b$ denote the number
of blocks obtained if $U_i$ = 0, conditional on $U_j$ , for $i \neq j$. The transition probability $p$, for the conditional probability of a change point at the position $i$ + 1, can be obtained from equation (\ref{equation1}) \cite{barry1993bayesian,erdman2007bcp}:  

\begin{equation}\label{equation1}
\begin{split}
\frac{p_i}{1-p_i} &= \frac{p(U_i=1|X,U_j,j \neq i)}{p(U_i=0|X,U_j,j \neq i)}  \\
&= \frac{\int_{0}^{\gamma} p^{b}(1-p)^{n-b-1} dp}{\int_{0}^{\gamma} p^{b-1}(1-p)^{n-b} dp} \frac{\int_{0}^{\lambda} \frac{w^{b/2}}{(W_1 + B_1w)^{(n-1)/2}}dw}{\int_{0}^{\lambda} \frac{w^{(b-1)/2}}{(W_0 + B_0w)^{(n-1)/2}}dw}
\end{split}
\end{equation}

Here $B_0, W_0$ and $B_1, W_1$ are the within and between block sums of squares obtained when $U_i$ = 0 (with change point at location $i$) and $U_i$ = 1 (without change point at location $i$), respectively. The two tuning parameters $\gamma$ and $\lambda$ can be calculated with MCMC. We use \textit{bcp} package in R \cite{erdman2007bcp} to implement the change point analysis. A similar approach has been utilized in a previous study to identify changes in driver's HR data in different roadway conditions \cite{tavakoli2021harmony}. The BCP output is a time series data of probability of change points.

\subsubsection{Gaze entropy}
Gaze entropy is a comprehensive measurement of visual scanning efficiency. The concept of entropy originates from information theory \cite{shannon1948mathematical}. It is only in the recent years that entropy has gained growing attention from researchers attempting to quantitatively examine gaze behavior in naturalistic settings. In according with Shannon’s
equation of \textit{entropy} and \textit{conditional entropy} \cite{shannon1948mathematical}, there are two types of gaze entropy measures: \textit{stationary gaze entropy} (SGE) and \textit{gaze transition entropy} (GTE) \cite{shiferaw2019review}. SGE measures overall predictability for fixation locations, which indicates the level of gaze dispersion during a given viewing period \cite{holmqvist2011eye}. The SGE is calculated using Shannon’s equation:
\begin{equation} \label{equation:SGE}
\begin{split}
H_{s}(x) = -\sum_{i=1}^{n}(p_i)log_2(p_i)
\end{split}
\end{equation}
Here $H_{s}(x)$ is the value of SGE for a sequence of data $x$ with length $n$, $i$ is the index for each individual state, $p_i$ is the proportion of each state within $x$, it is assumed that a fixation is an individual output of the gaze control system that makes spatial predictions regarding the location of subsequent fixations \cite{shiferaw2019review}. 

Gaze transition entropy (GTE) is conducted by applying the conditional entropy equation to 1st order Markov transitions of fixations with the following equation:
\begin{equation}
\begin{split}
H_{c}(x) = -\sum_{i=1}^{n}(p_i) \sum_{i=1}^{n}p(i,j) log_2 p(i,j)
\end{split}
\end{equation}
Here $H_{c}(x)$ is the value of GTE, $p_i$ is the stationary distribution, same as equation (\ref{equation:SGE}), and $p(i, j)$ is the probability of transitioning from i to j. GTE provides an overall estimation for the level of complexity or randomness in the pattern of visual scanning relative to overall spatial dispersion of gaze, where higher entropy suggests less predictability.

Specifically, to calculate the SGE and GTE, the visual field is divided into spatial bins of discrete state spaces to generate probability distributions. In this study, the fixation coordinates were divided into spatial bins of 100 × 100 pixel, followed a previous study \cite{shiferaw2018stationary}. To get the trend of gaze entropy, it is calculated in a rolling window of five seconds (600 data points in raw gaze data streams). We also apply the BCP methods to calculate the change points in the two gaze entropy values.

\section{Case Studies}
In this section, we present two case studies (one for bicyclists and one for pedestrians) from a pilot study of five participants to evaluate the proposed framework and highlight the importance of collecting physiological data, speed, and position data from participants. Using HR data, we demonstrate the relationship between the number of changes in HR data and the corresponding time frames with the changes in the environment (e.g., bicyclist arriving to an intersection) and/or the participant behaviors (e.g., when pedestrian is ready to cross the street). 
The tasks in the IVE are different for each user type (bicyclists and pedestrians) in the case study. The bicyclists were asked to cycle eastbound along the corridor, as indicated by Fig.\ref{fig:location}. The pedestrian task was to cross the street using the crosswalk at intersection 2 whenever they felt it was safe to do so. More details about the bicyclists experiment can be found in our previous study \cite{guo_robartes_angulo_chen_heydarian_2021}. Furthermore, we show how the multimodal dataset can be utilized to detect pedestrians and bicyclists’ state changes. The sample dataset is publicly accessible online \cite{guo_robartes_angulo_chen_heydarian_2021}. We first identify where abrupt changes happen in the HR readings and then identify the potential reasons behind the events that take place in a given time frame. To achieve this, the videos are manually annotated to identify event or behavioral changes among participants. Then the timestamps of these event/behavior changes, as well as other physiological responses are compared to the time that we observe HR change points for each participant. Through this, we can show whether the effect of HR changes is consistent across different groups of participants. 

The other physiological variables selected in the case study are: head movement direction, the position of the bicycle and pedestrian from Unity, gaze direction from eye tracker, and the gaze entropy and its BCP probability from the gaze direction.

\subsection{Bicycle Pilot Study}
In this experiment, after familiarization of the simulator and calibration for eye tracking and steering in a training scenario, the participants were asked to cycle eastbound in the simulated environment as indicated in Fig.\ref{fig:location}. 

\begin{figure}
\centering
\includegraphics[width=\textwidth]{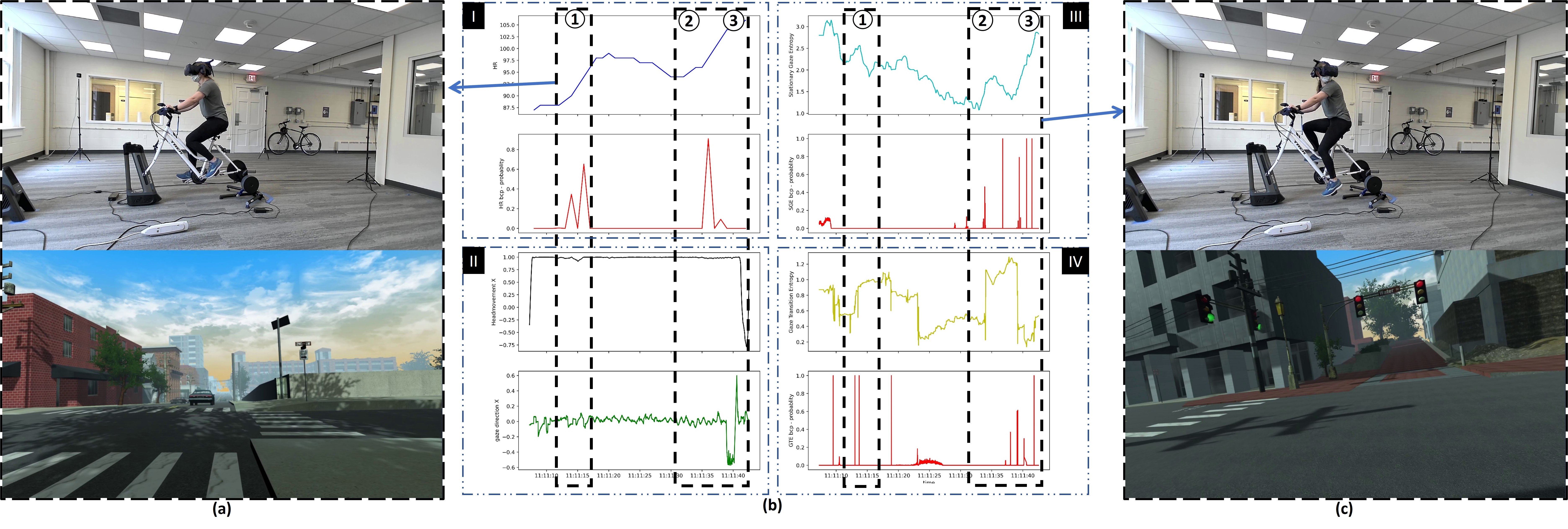}
\caption{Bicyclist’s HR change point analysis over time with other physiological responses. (a). Video snapshot of HR change point event 1. The bicyclist is approaching the intersection; (b). Visualizations of different physiological data: I. HR (blue) and HR BCP probability (red); II. horizontal head movement direction (black) and horizontal gaze direction (green); III. stationary gaze entropy (SGE) (cyan) and SGE BCP probability (red); IV. gaze transition entropy (GTE) (yellow) and GTE BCP probability (red); (c).Video snapshot of HR change point event 2. The bicyclist looked left behind to check if cars are approaching from behind.}
\label{fig:bike_case}
\end{figure}

Fig. \ref{fig:bike_case} shows one participant's physiological responses from the pilot bike experiment. Using BCP, we are able to detect the moments when the underlying distribution of HR data changes in a short period of time. Fig.\ref{fig:bike_case}b shows the overall time series of different physiological data. Fig.\ref{fig:bike_case}b.I. shows the HR (blue) and the probability of detected the change point events (red) during the whole experiment. In addition to the HR data, Fig.\ref{fig:bike_case}b.II. shows the the head movement x (black) and gaze direction x (green), the head movement in x-axis indicates the head facing direction from straight backward (-1) to straight forward (1). The gaze direction x indicates the gaze direction from left (-1) to right (1). Fig.\ref{fig:bike_case}b.III. shows the stationary gaze entropy (cyan) and BCP probability of SGE (red), Fig.\ref{fig:bike_case}b.IV. shows the gaze transition entropy (yellow) and BCP probability of SGE(red). 

Fig. \ref{fig:bike_case}a and c shows the corresponding screenshots for the two HR change points detected in Fig. \ref{fig:bike_case}b-I. The first change point happens when the participant is approaching the first intersection on the road which does not have any traffic signals; at this time, the participant is also being passed by a vehicle on the left (Fig. \ref{fig:bike_case}a). Meanwhile, the other physiological signals do not show abrupt changes except for minor peaks in GTE as shown in Fig. \ref{fig:bike_case}b-IV. The second HR change point takes place when the participant is approaching the third intersection where there is a traffic signal. While crossing the intersection, a looking-around behavior is also observed as shown in (Fig. \ref{fig:bike_case}c). As a result, we observe changes in both horizontal head and gaze direction (Fig. \ref{fig:bike_case}b-II), a larger variance in SGE and the change points detected from the SGE data points (Fig. \ref{fig:bike_case}b-III). Similarly, we observe higher variance and more change points in GTE (Fig. \ref{fig:bike_case}b-IV). Previous research suggests an increase in SGE associated with a
 higher GTE may reflect the influence of top-down interference on visual  scanning, which results in a greater dispersion of gaze \cite{shiferaw2019review}. In other words, increased SGE together with GTE indicates a higher visual or cognitive load in the experiment scenario for this participant. This case study indicates that the HR and gaze changes are sensitive to the environmental changes as well as the participant/bicyclist behaviors. It is also important to note that specific contextual factors (e.g., an intersection with or without a traffic signal) can trigger different physiological responses; therefore, it is important to collect and monitor different physiological data when conducting naturalistic or experimental studies of bicyclists. 

\begin{figure}
\centering
\includegraphics[width=0.95\textwidth]{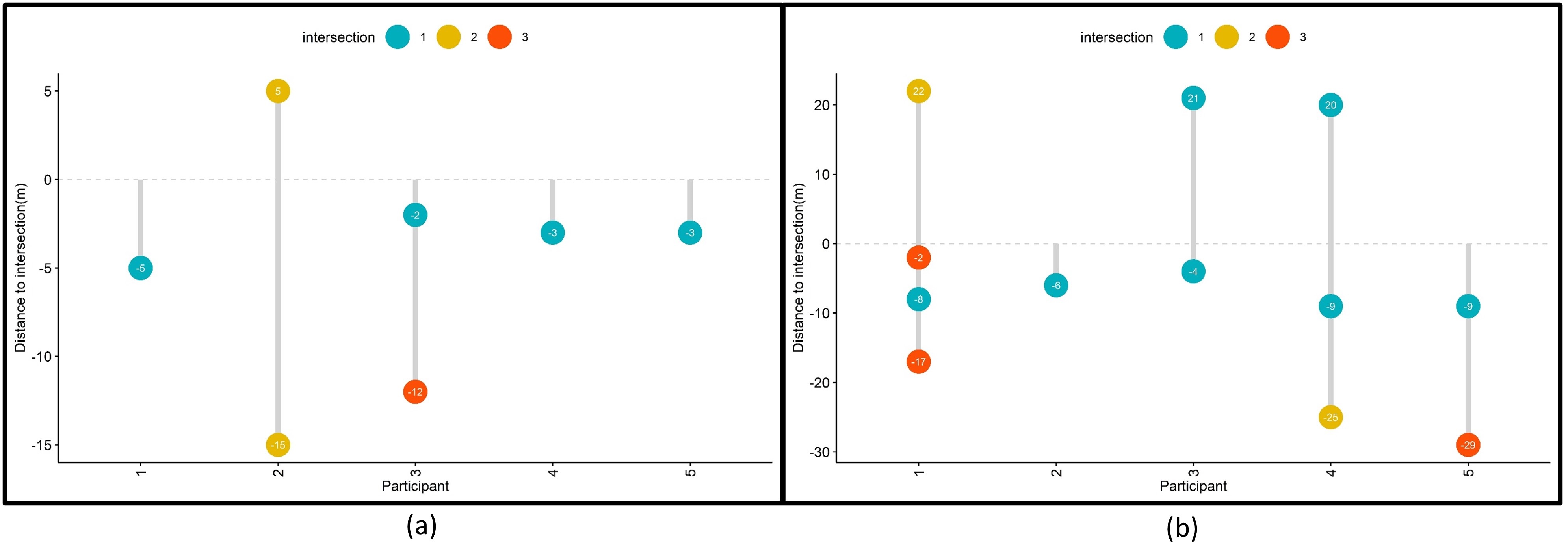}
\caption{Summary of where the HR (a) and gaze transition entropy (b) change point happen for bicycle experiment. Each dot indicates a HR/GTE change point. The x axis indicates the participant number, the y axis value is the distance to the intersection, where negative value indicates the the distance the change point happens before the participant arrives to the intersection. For example, for participant 2, has two HR change points at 15 meters prior intersection 2 as well as 5 meters after passing through that intersection (a). And participant 2 has one GTE change point 6 meters prior intersection 1}
\label{fig:bike HR change point}
\end{figure}

To find the reason behind each event, all the five participants' video recordings in the case study are manually analyzed. Fig.\ref{fig:bike HR change point} illustrates when the HR and gaze (use GTE as an example) change points happen for each participant. For HR, almost all the change points takes place when participants are approaching an intersection within a distance of 15 meters, except for Participant 2. When Participant 2 was passed by a vehicle in intersection 2 with a very close lateral proximity, the HR went up immediately (no other participant in the pilot study had a car pass by them as closely). For gaze transition entropy, the change point generally happens earlier than the HR change point, but follows a similar trend as the HR. Although the sample size is small, some of our observations from the case study include: 1) among the five participants in the pilot study, there are more HR/GTE change points prior to reaching the first intersection. As it is the first intersection in the experiment, participants may feel more stress than when approaching other intersections, as they became familiar with the environment. This implies that in the early portions of VR experiments, participants still need some time to get adjusted to the IVE environment, even after a training scenario before the actual experiment. 2) The change points prior to intersections 2 and 3 take place farther from the intersections compared to the change points detection prior to intersection 1. This could be explained by two possible factors: first, the road segment after intersection 1 has a 4\% downhill slope as shown in Fig. \ref{fig:location}d, where the participants' visibility of the road is limited until they get close or pass the intersection. Second, the roadway environment for intersections 2 and 3 are more complex. Intersection 2 is at the end of the downhill road segment and there is a lane shift after intersection 2, thus braking and right steering is needed before they enter intersection 2. Intersection 3, as indicated before, includes a traffic signal. Although participants are told the signal will always be green during the experiment, their physiological (HR and gaze entropy) data still showed a distinct response at this intersection. 

\subsection{Pedestrian Pilot Study}
The pedestrian pilot study is conducted at intersection 2 in the same IVEs \ref{fig:location} with the pedestrian simulator, where participants can walk freely as they would do in real life to cross a crosswalk. As explained before, the eastbound lane has randomly-generated vehicles with different gaps. At the beginning of the pilot study, participants are asked to wait until the first vehicle passes before they can cross using the crosswalk. Once the first car passes, whenever they feel safe, they may cross the road.

\begin{figure}
\centering
\includegraphics[width=\textwidth]{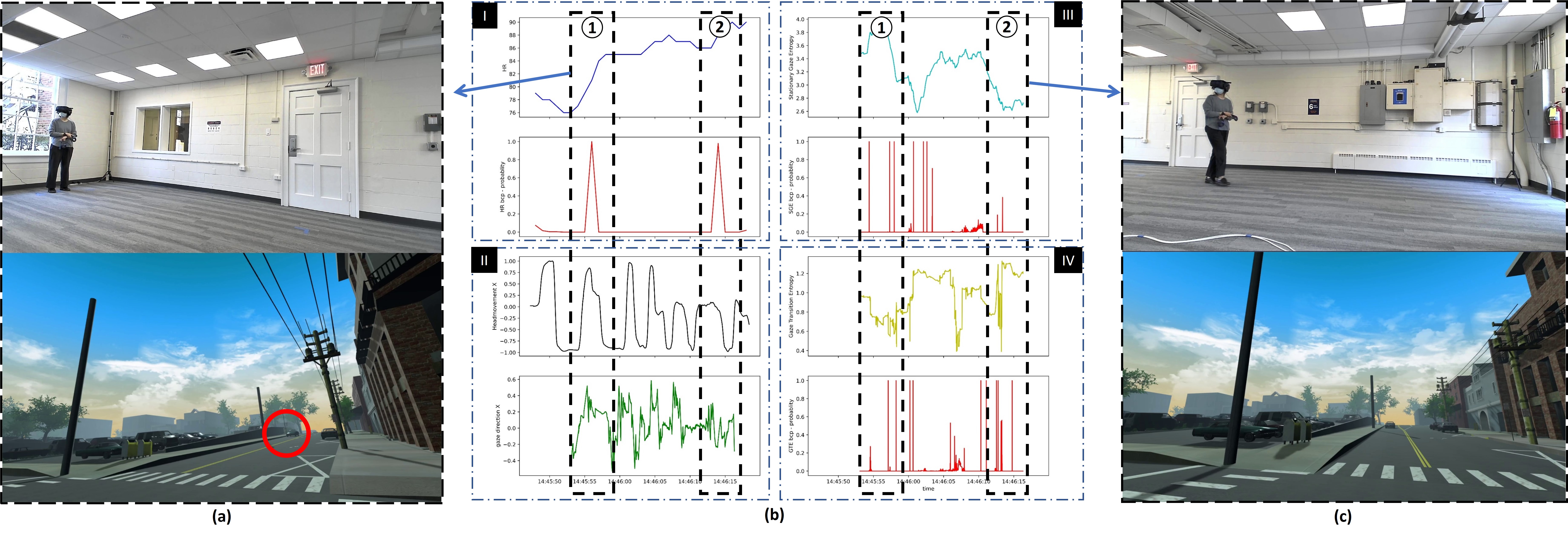}
\caption{Pedestrian’s HR change point analysis over time with other physiological responses. (a). Video snapshot of HR change point event 1. The pedestrian noticed the first approaching vehicle; (b). Visualizations of different physiological data: I. HR (blue) and HR BCP probability (red); II. horizontal head movement direction (black) and horizontal gaze direction (green); III. stationary gaze entropy (SGE) (cyan) and SGE BCP probability (red); IV. gaze transition entropy (GTE) (yellow) and GTE BCP probability (red); (c). Video snapshot of HR change point event 2. The pedestrian is crossing in the eastbound lane, just after taking a look at the approaching vehicle in the lane. }
\label{fig:ped_case}
\end{figure}

Similar to the bicycle case study, we extract the physiological data with the HR change point analysis results for one of the participants as shown in Fig.\ref{fig:ped_case}. The definition of the data is the same as the bicycle case study. The first change point happened when the pedestrian noticed the first approaching vehicle, as indicated by the red circle in Fig.\ref{fig:ped_case}a. A larger variance in SGE (Fig.\ref{fig:ped_case}b.III) and GTE (Fig.\ref{fig:ped_case}b.IV) is observed at the same time. An increase in SGE associated with lower GTE is likely indicative of distraction (such as the first approaching vehicle in this case). The second change point happened during the crossing in the eastbound lane, just after the participant looks at the approaching vehicle in the lane (Fig.\ref{fig:ped_case}c). During the change point event, only a larger variance in GTE (Fig.\ref{fig:ped_case}b.IV) is observed, while SGE remains at a low level (Fig.\ref{fig:ped_case}b.III). A reduction in SGE when GTE is increasing reflects top-down interference whereby the viewer focuses on specific items within the visual scene. In this case, the participant is looking straight to the other side of the road after a last look at the approaching vehicle in the lane, trying to cross the crosswalk quickly. In addition, after the pedestrian starts crossing, the range of horizontal head movement is smaller than before crossing (Fig.\ref{fig:ped_case}b.II). This indicates once they make the decision to cross, they will not observe the surroundings (e.g., incoming vehicles) as much as they do before crossing.

Table \ref{table:pedestrian HR change point} shows the video annotation details for pedestrian experiment. A total number of 7 HR change points are identified across the participants. There are three main categories within HR change points: two HR change points are detected when participants noticed the first approaching vehicle, two HR change points are identified when participants cross the crosswalk right after the first vehicle passes, and three HR change points are detected when participants are crossing in the vehicle-approaching (eastbound) lane. Similar to the bicycle pilot study, these change points correlate to the changes in the contextual setting, such as a vehicle is approaching a cross walk. These findings indicate why it is important to collect participants' physiological responses when conducting pedestrian studies. Although our preliminary findings show there exists a correlation between HR and gaze change points to the time that certain events takes place in the environment, analysis of a larger group of participants are needed to verify the findings. 

\begin{table}[h!]
\caption{Summary of pedestrian HR change point}
\label{table:pedestrian HR change point}
\centering
 \begin{tabular*}{\textwidth}{c c } 
  \hline
 Description of HR change point category & Included Participants   \\ [0.5ex]  \hline
 Noticed the first approaching vehicle in the initial position & 1,5    \\
 Start crossing after the first vehicle passed from the initial position & 3,4   \\
 Crossing in east bound lane after looking at the approaching vehicle &   1,2,5\\ 
 \hline
 \end{tabular*}
\end{table}

\section{Discussion and Conclusion} 
In this paper, we have developed a system architecture (ORCLSim) for VR simulators to capture physiological and behavioral changes in bicyclist and pedestrian studies. Specifically, the aim of this study is to determine (1) what metrics and set of information are needed to monitor bicyclists and pedestrians' behavioral changes, (2) what devices are available and how different hardware and software packages can be integrated in IVEs to conduct similar studies, (3) how the multi-modal data can be processed for observing the changes in physiological responses given different contextual settings, and (4) showcase how the proposed framework can be implemented by presenting two case studies for bicyclists and pedestrians. 

Previous studies on bicyclists and pedestrians' responses to changes in contextual settings highlight the advantages of controlled experimentation, especially in IVEs. In this paper, we demonstrated that it is important to track physiological metrics to better understand how different settings may impact the bicyclists and pedestrians. Additionally, we showcased the importance of gaze tracking and heart rate data in capturing behavioral response to how different events or roadway settings. These measurements may indicate the impact of the stress levels as well as the cognitive load on the participants. In the case study, our initial findings from the five participants indicate that physiological data is sensitive to road environment changes or real-time events, especially for the change in heart rate and gaze behavior. In the presented framework, we use Bayesian Change Point (BCP) detection method to detect abrupt changes in physiological data. First, we use the HR change point to identify any potential events, then the video annotation results can help to get a better understanding of the causes behind each event. The findings are further verified by two measurements of eye tracking data: stationary gaze entropy (SGE) and gaze transition entropy (GTE). The dynamical changes in the eye tracking data also support the observations from the video annotation. For the presented bicycle case study, most change points happen prior to the intersections, while the eye tracking change points usually happened earlier than the HR change points. The increased SGE and GTE along with abrupt changes in HR indicates where the participants feel stressful in the environment, which are observed to be at the beginning of the experiment and when participants reach the intersection with a traffic signal. The physiological changes in the pedestrian case study are indicative of critical behavior during crossing, such as observing the first approaching vehicle or the moment before crossing. Furthermore, the differences between individual participants' physiological responses also emphasize the importance of building personalized models for different groups of people. Although these preliminary findings are promising, we need to further examine whether these changes points are observed when the number of participants are increased for both case studies.  

We have open-sourced the system set-up document, code example, and sample dataset for the research community. The integration between the presented devices and software platforms along with the data processing method provide the foundation to support IVE experimental studies where we can identify the impact of different roadway designs on bicyclist and pedestrian behavioral and psychological changes. This presented system architecture can be used to study bicyclists and pedestrians' behaviors that may be affected by their perception ability and cognitive state, which may also be influenced by different road design conditions. Furthermore, it makes the development of a VR simulator simpler and more robust since many of the modules are flexible and scalable to  different systems and improvements. For example, the smartwatch system can be replaced by more recent and advanced wearable devices that can collect different data streams; or the video recording systems can integrate more event or activity detection through computer vision based techniques.

\subsection{Limitations and Future work}
While useful in addressing many of the gaps in virtual simulation research, IVEs have some limitations. Many of the reports included in Table \ref{bicycle literature table} and Table \ref{pedestrian literature table} indicate that portions of their subject pool’s data had to be disregarded due to the motion sickness participants experienced while in VR. 

Furthermore, the removal of risk within the IVE may also be perceptible to participants – IVE experimentation relies heavily on subject immersion and while the environment may look, feel, and behave real, the knowledge that one is still in a risk-free virtual space where physical injury is not possible still exists. It is up to the realism of the IVE to suspend a subject’s disbelief in the environment sufficiently to overcome this knowledge, which varies from person to person. Additionally, a subject's familiarity with VR technology could have an impact on their behavior and how they perceive the IVE - someone who has played multiple games within VR may be aware that if they were to collide with an object, they would not physically be affected.

With our system architecture ORCLSim, future IVE research can apply any physiological data collection modules to their own IVE simulators to study the vulnerable road users' behaviors, perception abilities and cognitive states in different contextual settings. Additionally, more physiological responses may be included in the system with off-the-shelf sensors, such as Electrodermal activity (EDA) and Electromyography (EMG). Future IVE can be improved to increase immersion and tackle more complex research problems. A feature that would greatly improve the ease of development of an IVE would be the integration of development platforms, such as Unity or Unreal Engine, and commercially available transportation simulation software, such as Synchro and Vissim. Through such integration, roadway segments, objects, vehicles, vehicle behaviors, and traffic networks could be  simulated more realistically and robustly within IVE. Furthermore, more robust platforms for integrating multiple users into IVEs would vastly improve immersion and realism. Instead of modelling vehicles to interact with bicyclists and pedestrians, having another subject controlling a vehicle via a driving simulator and interacting with a bicyclist and a pedestrian would provide valuable insights into the interactions between road users unlike anything that has been done before.

\section{Author contributions}
\textbf{Xiang Guo}: Methodology, software, validation, formal analysis, investigation, data curation, writing - original draft, visualization. \textbf{Austin Angulo}: Methodology, software, investigation, resources, data curation, writing- original draft preparation. \textbf{Erin Robartes}: Software, investigation, resources, writing- original draft preparation.\textbf{ T. Donna Chen}: conceptualization, writing - review \& editing, supervision, project administration, funding acquisition. \textbf{Arsalan Heydarian}: Conceptualization, validation, writing - review \& editing, supervision, project administration, funding acquisition.

\pagebreak
%

%
%
\bibliography{ascexmpl-new}

\begin{thebibliography}{}

\bibitem[\protect\citeauthoryear{}{Abadi and
  Hurwitz}{2018}]{abadi2018bicyclist}
Abadi, M.~G. and Hurwitz, D.~S. (2018).
\newblock ``Bicyclist’s perceived level of comfort in dense urban
  environments: How do ambient traffic, engineering treatments, and bicyclist
  characteristics relate?.''\ {\em Sustainable cities and society}, 40,
  101--109.

\bibitem[\protect\citeauthoryear{}{Abadi et~al.\@}{2019}]{abadi2019factors}
Abadi, M.~G., Hurwitz, D.~S., Sheth, M., McCormack, E., and Goodchild, A.
  (2019).
\newblock ``Factors impacting bicyclist lateral position and velocity in
  proximity to commercial vehicle loading zones: Application of a bicycling
  simulator.''\ {\em Accident Analysis \& Prevention}, 125, 29--39.

\bibitem[\protect\citeauthoryear{}{Adami
  et~al.\@}{2021}]{adami2021effectiveness}
Adami, P., Rodrigues, P.~B., Woods, P.~J., Becerik-Gerber, B., Soibelman, L.,
  Copur-Gencturk, Y., and Lucas, G. (2021).
\newblock ``Effectiveness of vr-based training on improving construction
  workers’ knowledge, skills, and safety behavior in robotic
  teleoperation.''\ {\em Advanced Engineering Informatics}, 50, 101431.

\bibitem[\protect\citeauthoryear{}{Akbar et~al.\@}{2017}]{akbar2017three}
Akbar, I.~A., Rumagit, A.~M., Utsunomiya, M., Morie, T., and Igasaki, T.
  (2017).
\newblock ``Three drowsiness categories assessment by electroencephalogram in
  driving simulator environment.''\ {\em 2017 39th Annual International
  Conference of the IEEE Engineering in Medicine and Biology Society (EMBC)},
  IEEE,  2904--2907.

\bibitem[\protect\citeauthoryear{}{Awada et~al.\@}{2021}]{awada2021integrated}
Awada, M., Zhu, R., Becerik-Gerber, B., Lucas, G., and Southers, E. (2021).
\newblock ``An integrated emotional and physiological assessment for vr-based
  active shooter incident experiments.''\ {\em Advanced Engineering
  Informatics}, 47, 101227.

\bibitem[\protect\citeauthoryear{}{Ayres et~al.\@}{2015}]{ayres2015bicyclist}
Ayres, T.~J., Kelkar, R., Kubose, T., and Shekhawat, V. (2015).
\newblock ``Bicyclist behavior at stop signs.''\ {\em Proceedings of the Human
  Factors and Ergonomics Society Annual Meeting}, Vol.~59, SAGE Publications
  Sage CA: Los Angeles, CA,  1616--1620.

\bibitem[\protect\citeauthoryear{}{Baee et~al.\@}{2021}]{baee2021medirl}
Baee, S., Pakdamanian, E., Kim, I., Feng, L., Ordonez, V., and Barnes, L.
  (2021).
\newblock ``Medirl: Predicting the visual attention of drivers via maximum
  entropy deep inverse reinforcement learning.''\ {\em Proceedings of the
  IEEE/CVF International Conference on Computer Vision},  13178--13188.

\bibitem[\protect\citeauthoryear{}{Banton
  et~al.\@}{2005}]{banton2005perception}
Banton, T., Stefanucci, J., Durgin, F., Fass, A., and Proffitt, D. (2005).
\newblock ``The perception of walking speed in a virtual environment.''\ {\em
  Presence}, 14(4), 394--406.

\bibitem[\protect\citeauthoryear{}{Barry and
  Hartigan}{1993}]{barry1993bayesian}
Barry, D. and Hartigan, J.~A. (1993).
\newblock ``A bayesian analysis for change point problems.''\ {\em Journal of
  the American Statistical Association}, 88(421), 309--319.

\bibitem[\protect\citeauthoryear{}{Bernhard
  et~al.\@}{2008}]{bernhard2008bimodal}
Bernhard, M., Grosse, K., and Wimmer, M. (2008).
\newblock ``Bimodal task-facilitation in a virtual traffic scenario through
  spatialized sound rendering.''\ {\em ACM Transactions on Applied Perception
  (TAP)}, 8(4), 1--22.

\bibitem[\protect\citeauthoryear{}{Bhagavathula
  et~al.\@}{2018}]{bhagavathula2018reality}
Bhagavathula, R., Williams, B., Owens, J., and Gibbons, R. (2018).
\newblock ``The reality of virtual reality: A comparison of pedestrian behavior
  in real and virtual environments.''\ {\em Proceedings of the Human Factors
  and Ergonomics Society Annual Meeting}, Vol.~62, SAGE Publications Sage CA:
  Los Angeles, CA,  2056--2060.

\bibitem[\protect\citeauthoryear{}{Bogacz et~al.\@}{2021}]{bogacz2021modelling}
Bogacz, M., Hess, S., Calastri, C., Choudhury, C.~F., Mushtaq, F., Awais, M.,
  Nazemi, M., van Eggermond, M.~A., and Erath, A. (2021).
\newblock ``Modelling risk perception using a dynamic hybrid choice model and
  brain-imaging data: Application to virtual reality cycling.''\ {\em
  Transportation Research Part C: Emerging Technologies}, 133, 103435.

\bibitem[\protect\citeauthoryear{}{Boukhechba and
  Barnes}{2020}]{boukhechba2020swear}
Boukhechba, M. and Barnes, L.~E. (2020).
\newblock ``Swear: Sensing using wearables. generalized human crowdsensing on
  smartwatches.''\ {\em International Conference on Applied Human Factors and
  Ergonomics}, Springer,  510--516.

\bibitem[\protect\citeauthoryear{}{Brown
  et~al.\@}{2017}]{brown2017investigation}
Brown, H., Sun, C., and Qing, Z. (2017).
\newblock ``Investigation of alternative bicycle pavement markings with the use
  of a bicycle simulator.''\ {\em Transportation research record}, 2662(1),
  143--151.

\bibitem[\protect\citeauthoryear{}{Cao et~al.\@}{2017}]{cao2017realtime}
Cao, Z., Simon, T., Wei, S.-E., and Sheikh, Y. (2017).
\newblock ``Realtime multi-person 2d pose estimation using part affinity
  fields.''\ {\em Proceedings of the IEEE conference on computer vision and
  pattern recognition},  7291--7299.

\bibitem[\protect\citeauthoryear{}{Cavallo et~al.\@}{2019}]{cavallo2019street}
Cavallo, V., Dommes, A., Dang, N.-T., and Vienne, F. (2019).
\newblock ``A street-crossing simulator for studying and training
  pedestrians.''\ {\em Transportation research part F: traffic psychology and
  behaviour}, 61, 217--228.

\bibitem[\protect\citeauthoryear{}{Chapman
  et~al.\@}{2008}]{chapman2008patterns}
Chapman, A.~R., Vicenzino, B., Blanch, P., and Hodges, P.~W. (2008).
\newblock ``Patterns of leg muscle recruitment vary between novice and highly
  trained cyclists.''\ {\em Journal of Electromyography and Kinesiology},
  18(3), 359--371.

\bibitem[\protect\citeauthoryear{}{Chaurand and
  Delhomme}{2013}]{chaurand2013cyclists}
Chaurand, N. and Delhomme, P. (2013).
\newblock ``Cyclists and drivers in road interactions: A comparison of
  perceived crash risk.''\ {\em Accident Analysis \& Prevention}, 50,
  1176--1184.

\bibitem[\protect\citeauthoryear{}{Chen and Shen}{2016}]{chen2016built}
Chen, P. and Shen, Q. (2016).
\newblock ``Built environment effects on cyclist injury severity in
  automobile-involved bicycle crashes.''\ {\em Accident Analysis \&
  Prevention}, 86, 239--246.

\bibitem[\protect\citeauthoryear{}{Ch{\'\i}as~Navarro
  et~al.\@}{2019}]{chias20193d}
Ch{\'\i}as~Navarro, P., Abad~Balboa, T., Miguel~S{\'a}nchez, M.~d.,
  Garc{\'\i}a-Rosales Gonz{\'a}lez-Fierro, G., Echeverr{\'\i}a~Valiente, E.,
  et~al.\@ (2019).
\newblock ``3d modelling and virtual reality applied to complex architectures:
  an application to hospitals' design.

\bibitem[\protect\citeauthoryear{}{Cloutier et~al.\@}{2017}]{cloutier2017outta}
Cloutier, M.-S., Lachapelle, U., d’Amours Ouellet, A.-A., Bergeron, J., Lord,
  S., and Torres, J. (2017).
\newblock ``“outta my way!” individual and environmental correlates of
  interactions between pedestrians and vehicles during street crossings.''\
  {\em Accident Analysis \& Prevention}, 104, 36--45.

\bibitem[\protect\citeauthoryear{}{Cobb et~al.\@}{2021}]{cobb2021bicyclists}
Cobb, D.~P., Jashami, H., and Hurwitz, D.~S. (2021).
\newblock ``Bicyclists’ behavioral and physiological responses to varying
  roadway conditions and bicycle infrastructure.''\ {\em Transportation
  Research Part F: Traffic Psychology and Behaviour}, 80, 172--188.

\bibitem[\protect\citeauthoryear{}{Correa et~al.\@}{2016}]{correa2016indoor}
Correa, A., Llado, M.~B., Morell, A., and Vicario, J.~L. (2016).
\newblock ``Indoor pedestrian tracking by on-body multiple receivers.''\ {\em
  IEEE Sensors Journal}, 16(8), 2545--2553.

\bibitem[\protect\citeauthoryear{}{Deb et~al.\@}{2017}]{deb2017efficacy}
Deb, S., Carruth, D.~W., Sween, R., Strawderman, L., and Garrison, T.~M.
  (2017).
\newblock ``Efficacy of virtual reality in pedestrian safety research.''\ {\em
  Applied ergonomics}, 65, 449--460.

\bibitem[\protect\citeauthoryear{}{Deb et~al.\@}{2018}]{deb2018investigating}
Deb, S., Strawderman, L.~J., and Carruth, D.~W. (2018).
\newblock ``Investigating pedestrian suggestions for external features on fully
  autonomous vehicles: A virtual reality experiment.''\ {\em Transportation
  research part F: traffic psychology and behaviour}, 59, 135--149.

\bibitem[\protect\citeauthoryear{}{Erdman and Emerson}{2007}]{erdman2007bcp}
Erdman, C. and Emerson, J.~W. (2007).
\newblock ``bcp: an r package for performing a bayesian analysis of change
  point problems.''\ {\em Journal of Statistical Software}, 23(1), 1--13.

\bibitem[\protect\citeauthoryear{}{Ergan et~al.\@}{2019}]{ergan2019quantifying}
Ergan, S., Radwan, A., Zou, Z., Tseng, H.-a., and Han, X. (2019).
\newblock ``Quantifying human experience in architectural spaces with
  integrated virtual reality and body sensor networks.''\ {\em Journal of
  Computing in Civil Engineering}, 33(2), 04018062.

\bibitem[\protect\citeauthoryear{}{Eudave and
  Valencia}{2017}]{eudave2017physiological}
Eudave, L. and Valencia, M. (2017).
\newblock ``Physiological response while driving in an immersive virtual
  environment.''\ {\em 2017 IEEE 14th International Conference on Wearable and
  Implantable Body Sensor Networks (BSN)}, IEEE,  145--148.

\bibitem[\protect\citeauthoryear{}{Exler et~al.\@}{2016}]{exler2016wearable}
Exler, A., Schankin, A., Klebsattel, C., and Beigl, M. (2016).
\newblock ``A wearable system for mood assessment considering smartphone
  features and data from mobile ecgs.''\ {\em Proceedings of the 2016 ACM
  international joint conference on pervasive and ubiquitous computing:
  Adjunct},  1153--1161.

\bibitem[\protect\citeauthoryear{}{Farooq et~al.\@}{2018}]{farooq2018virtual}
Farooq, B., Cherchi, E., and Sobhani, A. (2018).
\newblock ``Virtual immersive reality for stated preference travel behavior
  experiments: A case study of autonomous vehicles on urban roads.''\ {\em
  Transportation research record}, 2672(50), 35--45.

\bibitem[\protect\citeauthoryear{}{Fitch and
  Handy}{2018}]{fitch2018relationship}
Fitch, D.~T. and Handy, S.~L. (2018).
\newblock ``The relationship between experienced and imagined bicycling comfort
  and safety.''\ {\em Transportation research record}, 2672(36), 116--124.

\bibitem[\protect\citeauthoryear{}{Fitch
  et~al.\@}{2020}]{fitch2020psychological}
Fitch, D.~T., Sharpnack, J., and Handy, S.~L. (2020).
\newblock ``Psychological stress of bicycling with traffic: examining heart
  rate variability of bicyclists in natural urban environments.''\ {\em
  Transportation research part F: traffic psychology and behaviour}, 70,
  81--97.

\bibitem[\protect\citeauthoryear{}{Francisco
  et~al.\@}{2018}]{francisco2018occupant}
Francisco, A., Truong, H., Khosrowpour, A., Taylor, J.~E., and Mohammadi, N.
  (2018).
\newblock ``Occupant perceptions of building information model-based energy
  visualizations in eco-feedback systems.''\ {\em Applied Energy}, 221,
  220--228.

\bibitem[\protect\citeauthoryear{}{Guo}{2021}]{xiangwebsite}
Guo, X. (2021).
\newblock ``Orcl vr eyetracking set up, $<$https://git.io/JK3hq$>$.

\bibitem[\protect\citeauthoryear{}{Guo et~al.\@}{2019}]{guo2019will}
Guo, X., Cui, L., Park, B., Ding, W., Lockhart, M., and Kim, I. (2019).
\newblock ``How will humans cut through automated vehicle platoons in mixed
  traffic environments? a simulation study of drivers’ gaze behaviors based
  on the dynamic areas of interest.''\ {\em Systems Engineering in Context},
  Springer,  691--701.

\bibitem[\protect\citeauthoryear{}{Guo
  et~al.\@}{2021}]{guo_robartes_angulo_chen_heydarian_2021}
Guo, X., Robartes, E.~M., Angulo, A., Chen, T.~D., and Heydarian, A. (2021).
\newblock ``Benchmarking the use of immersive virtual bike simulators for
  understanding cyclist behaviors, $<$engrxiv.org/mrxgh$>$\ (Jul).

\bibitem[\protect\citeauthoryear{}{Haufe et~al.\@}{2011}]{haufe2011eeg}
Haufe, S., Treder, M.~S., Gugler, M.~F., Sagebaum, M., Curio, G., and
  Blankertz, B. (2011).
\newblock ``Eeg potentials predict upcoming emergency brakings during simulated
  driving.''\ {\em Journal of neural engineering}, 8(5), 056001.

\bibitem[\protect\citeauthoryear{}{Heydarian and
  Becerik-Gerber}{2017}]{heydarian2017use}
Heydarian, A. and Becerik-Gerber, B. (2017).
\newblock ``Use of immersive virtual environments for occupant behaviour
  monitoring and data collection.''\ {\em Journal of Building Performance
  Simulation}, 10(5-6), 484--498.

\bibitem[\protect\citeauthoryear{}{Heydarian
  et~al.\@}{2015}]{heydarian2015immersive}
Heydarian, A., Carneiro, J.~P., Gerber, D., Becerik-Gerber, B., Hayes, T., and
  Wood, W. (2015).
\newblock ``Immersive virtual environments versus physical built environments:
  A benchmarking study for building design and user-built environment
  explorations.''\ {\em Automation in Construction}, 54, 116--126.

\bibitem[\protect\citeauthoryear{}{Holmqvist et~al.\@}{2011}]{holmqvist2011eye}
Holmqvist, K., Nystr{\"o}m, M., Andersson, R., Dewhurst, R., Jarodzka, H., and
  Van~de Weijer, J. (2011).
\newblock {\em Eye tracking: A comprehensive guide to methods and measures}.
\newblock OUP Oxford.

\bibitem[\protect\citeauthoryear{}{Hoover et~al.\@}{2012}]{hoover2012real}
Hoover, A., Singh, A., Fishel-Brown, S., and Muth, E. (2012).
\newblock ``Real-time detection of workload changes using heart rate
  variability.''\ {\em Biomedical Signal Processing and Control}, 7(4),
  333--341.

\bibitem[\protect\citeauthoryear{}{Iryo-Asano
  et~al.\@}{2018}]{iryo2018applicability}
Iryo-Asano, M., Hasegawa, Y., and Dias, C. (2018).
\newblock ``Applicability of virtual reality systems for evaluating
  pedestrians’ perception and behavior.''\ {\em Transportation research
  procedia}, 34, 67--74.

\bibitem[\protect\citeauthoryear{}{Jiang et~al.\@}{2017}]{jiang2017influence}
Jiang, Y., O'Neal, E.~E., Franzen, L., Yon, J.~P., Plumert, J.~M., and Kearney,
  J.~K. (2017).
\newblock ``The influence of stereoscopic image display on pedestrian road
  crossing in a large-screen virtual environment.''\ {\em Proceedings of the
  ACM Symposium on Applied Perception},  1--4.

\bibitem[\protect\citeauthoryear{}{Jorna}{1992}]{jorna1992spectral}
Jorna, P.~G. (1992).
\newblock ``Spectral analysis of heart rate and psychological state: A review
  of its validity as a workload index.''\ {\em Biological psychology}, 34(2-3),
  237--257.

\bibitem[\protect\citeauthoryear{}{Keler et~al.\@}{2018}]{keler2018bicycle}
Keler, A., Kaths, J., Chucholowski, F., Chucholowski, M., Grigoropoulos, G.,
  Spangler, M., Kaths, H., and Busch, F. (2018).
\newblock ``A bicycle simulator for experiencing microscopic traffic flow
  simulation in urban environments.''\ {\em 2018 21st International Conference
  on Intelligent Transportation Systems (ITSC)}, IEEE,  3020--3023.

\bibitem[\protect\citeauthoryear{}{Kim et~al.\@}{2019}]{kim2019saliency}
Kim, J., Yadav, M., Ahn, C.~R., and Chaspari, T. (2019).
\newblock ``Saliency detection analysis of pedestrians’ physiological
  responses to assess adverse built environment features.''\ {\em Computing in
  Civil Engineering 2019: Smart Cities, Sustainability, and Resilience},
  American Society of Civil Engineers Reston, VA,  90--97.

\bibitem[\protect\citeauthoryear{}{Kim et~al.\@}{2020}]{kim2020environmental}
Kim, J., Yadav, M., Chaspari, T., and Ahn, C.~R. (2020).
\newblock ``Environmental distress and physiological signals: Examination of
  the saliency detection method.''\ {\em Journal of Computing in Civil
  Engineering}, 34(6), 04020046.

\bibitem[\protect\citeauthoryear{}{Kwigizile
  et~al.\@}{2017}]{kwigizile2017real}
Kwigizile, V., Oh, J.-S., Ikonomov, P., Hasan, R., Villalobos, C.~G., Kurdi,
  A.~H., Shaw, A., et~al.\@ (2017).
\newblock ``Real time bicycle simulation study of bicyclists’ behaviors and
  their implication on safety.''\ {\em Report no.}, Western Michigan
  University. Transportation Research Center for Livable~….

\bibitem[\protect\citeauthoryear{}{Kwon et~al.\@}{2001}]{kwon2001kaist}
Kwon, D.-S., Yang, G.-H., Lee, C.-W., Shin, J.-C., Park, Y., Jung, B., Lee,
  D.~Y., Lee, K., Han, S.-H., Yoo, B.-H., et~al.\@ (2001).
\newblock ``Kaist interactive bicycle simulator.''\ {\em Proceedings 2001 ICRA.
  IEEE International Conference on Robotics and Automation (Cat. No.
  01CH37164)}, Vol.~3, IEEE,  2313--2318.

\bibitem[\protect\citeauthoryear{}{Lee et~al.\@}{2020}]{lee2020wearable}
Lee, G., Choi, B., Jebelli, H., Ahn, C.~R., and Lee, S. (2020).
\newblock ``Wearable biosensor and collective sensing--based approach for
  detecting older adults’ environmental barriers.''\ {\em Journal of
  Computing in Civil Engineering}, 34(2), 04020002.

\bibitem[\protect\citeauthoryear{}{Lee et~al.\@}{2017}]{lee2017description}
Lee, O., Dialynas, G., De~Winter, J., Happee, R., and Schwab, A. (2017).
\newblock ``Description of a model based bicycle simulator.''\ {\em Proceedings
  of the 6th Annual International Cycling Safety Conference}.

\bibitem[\protect\citeauthoryear{}{Maillot
  et~al.\@}{2017}]{maillot2017training}
Maillot, P., Dommes, A., Dang, N.-T., and Vienne, F. (2017).
\newblock ``Training the elderly in pedestrian safety: transfer effect between
  two virtual reality simulation devices.''\ {\em Accident Analysis \&
  Prevention}, 99, 161--170.

\bibitem[\protect\citeauthoryear{}{Mallaro
  et~al.\@}{2017}]{mallaro2017comparison}
Mallaro, S., Rahimian, P., O'Neal, E.~E., Plumert, J.~M., and Kearney, J.~K.
  (2017).
\newblock ``A comparison of head-mounted displays vs. large-screen displays for
  an interactive pedestrian simulator.''\ {\em Proceedings of the 23rd ACM
  Symposium on Virtual Reality Software and Technology},  1--4.

\bibitem[\protect\citeauthoryear{}{Montuwy
  et~al.\@}{2016}]{montuwy2016adapting}
Montuwy, A., C{\oe}ugnet, S., and Dommes, A. (2016).
\newblock ``Adapting a pedestrian navigation simulator to the elderly.''\ {\em
  Proceedings of the European Conference on Cognitive Ergonomics},  1--2.

\bibitem[\protect\citeauthoryear{}{National Highway Traffic
  Safety~Administration}{2020a}]{national2020fatality}
National Highway Traffic Safety~Administration, N. (2020a).
\newblock ``Fatality analysisi reporting system encyclopedia.''\ {\em
  http://www-fars. nhtsa. dot. gov/Main/index. aspx}.

\bibitem[\protect\citeauthoryear{}{National Highway Traffic
  Safety~Administration}{2020b}]{NHTSA2020}
National Highway Traffic Safety~Administration, N. (2020b).
\newblock ``Overview of motor vehicle crashes in 2019.

\bibitem[\protect\citeauthoryear{}{Nazemi et~al.\@}{2018}]{nazemi2018speed}
Nazemi, M., van Eggermond, M.~A., Erath, A., Schaffner, D., Joos, M., and
  Axhausen, K.~W. (2018).
\newblock ``Speed and space perception in virtual reality for bicycle
  research.''\ {\em 7th International Cycling Safety Conference (ICSC 2018)}.

\bibitem[\protect\citeauthoryear{}{Neider
  et~al.\@}{2010}]{neider2010pedestrians}
Neider, M.~B., McCarley, J.~S., Crowell, J.~A., Kaczmarski, H., and Kramer,
  A.~F. (2010).
\newblock ``Pedestrians, vehicles, and cell phones.''\ {\em Accident Analysis
  \& Prevention}, 42(2), 589--594.

\bibitem[\protect\citeauthoryear{}{Nikolas et~al.\@}{2016}]{nikolas2016risky}
Nikolas, M.~A., Elmore, A.~L., Franzen, L., O'Neal, E., Kearney, J.~K., and
  Plumert, J.~M. (2016).
\newblock ``Risky bicycling behavior among youth with and without
  attention-deficit hyperactivity disorder.''\ {\em Journal of child psychology
  and psychiatry}, 57(2), 141--148.

\bibitem[\protect\citeauthoryear{}{Noghabaei and
  Han}{2020}]{noghabaei2020hazard}
Noghabaei, M. and Han, K. (2020).
\newblock ``Hazard recognition in an immersive virtual environment: Framework
  for the simultaneous analysis of visual search and eeg patterns.''\ {\em
  Construction Research Congress 2020: Computer Applications}, American Society
  of Civil Engineers Reston, VA,  934--943.

\bibitem[\protect\citeauthoryear{}{Noghabaei
  et~al.\@}{2021}]{noghabaei2021feasibility}
Noghabaei, M., Han, K., and Albert, A. (2021).
\newblock ``Feasibility study to identify brain activity and eye-tracking
  features for assessing hazard recognition using consumer-grade wearables in
  an immersive virtual environment.''\ {\em Journal of Construction Engineering
  and Management}, 147(9), 04021104.

\bibitem[\protect\citeauthoryear{}{O’Hern et~al.\@}{2017}]{o2017validation}
O’Hern, S., Oxley, J., and Stevenson, M. (2017).
\newblock ``Validation of a bicycle simulator for road safety research.''\ {\em
  Accident Analysis \& Prevention}, 100, 53--58.

\bibitem[\protect\citeauthoryear{}{Parkin et~al.\@}{2007}]{parkin2007models}
Parkin, J., Wardman, M., and Page, M. (2007).
\newblock ``Models of perceived cycling risk and route acceptability.''\ {\em
  Accident Analysis \& Prevention}, 39(2), 364--371.

\bibitem[\protect\citeauthoryear{}{Poulos et~al.\@}{2012}]{poulos2012exposure}
Poulos, R.~G., Hatfield, J., Rissel, C., Grzebieta, R., and McIntosh, A.~S.
  (2012).
\newblock ``Exposure-based cycling crash, near miss and injury rates: The safer
  cycling prospective cohort study protocol.''\ {\em Injury prevention}, 18(1),
  e1--e1.

\bibitem[\protect\citeauthoryear{}{Ridel et~al.\@}{2018}]{ridel2018literature}
Ridel, D., Rehder, E., Lauer, M., Stiller, C., and Wolf, D. (2018).
\newblock ``A literature review on the prediction of pedestrian behavior in
  urban scenarios.''\ {\em 2018 21st International Conference on Intelligent
  Transportation Systems (ITSC)}, IEEE,  3105--3112.

\bibitem[\protect\citeauthoryear{}{Robartes and Chen}{2018}]{robartes2018crash}
Robartes, E. and Chen, T.~D. (2018).
\newblock ``Crash histories, safety perceptions, and attitudes among virginia
  bicyclists.''\ {\em Journal of safety research}, 67, 189--196.

\bibitem[\protect\citeauthoryear{}{Rodriguez-Valencia
  et~al.\@}{2021}]{rodriguez2021towards}
Rodriguez-Valencia, A., Vallejo-Borda, J.~A., Barrero, G.~A., and
  Ortiz-Ramirez, H.~A. (2021).
\newblock ``Towards an enriched framework of service evaluation for pedestrian
  and bicyclist infrastructure: acknowledging the power of users’
  perceptions.''\ {\em Transportation},  1--24.

\bibitem[\protect\citeauthoryear{}{Rupi and Krizek}{2019}]{rupi2019visual}
Rupi, F. and Krizek, K.~J. (2019).
\newblock ``Visual eye gaze while cycling: Analyzing eye tracking at signalized
  intersections in urban conditions.''\ {\em Sustainability}, 11(21), 6089.

\bibitem[\protect\citeauthoryear{}{Rybarczyk
  et~al.\@}{2020}]{rybarczyk2020physiological}
Rybarczyk, G., Ozbil, A., Andresen, E., and Hayes, Z. (2020).
\newblock ``Physiological responses to urban design during bicycling: A
  naturalistic investigation.''\ {\em Transportation research part F: traffic
  psychology and behaviour}, 68, 79--93.

\bibitem[\protect\citeauthoryear{}{Schneider and
  Bengler}{2020}]{schneider2020virtually}
Schneider, S. and Bengler, K. (2020).
\newblock ``Virtually the same? analysing pedestrian behaviour by means of
  virtual reality.''\ {\em Transportation research part F: traffic psychology
  and behaviour}, 68, 231--256.

\bibitem[\protect\citeauthoryear{}{Schwebel
  et~al.\@}{2017a}]{schwebel2017experiential}
Schwebel, D.~C., McClure, L.~A., and Porter, B.~E. (2017a).
\newblock ``Experiential exposure to texting and walking in virtual reality: A
  randomized trial to reduce distracted pedestrian behavior.''\ {\em Accident
  Analysis \& Prevention}, 102, 116--122.

\bibitem[\protect\citeauthoryear{}{Schwebel
  et~al.\@}{2017b}]{schwebel2017using}
Schwebel, D.~C., Severson, J., and He, Y. (2017b).
\newblock ``Using smartphone technology to deliver a virtual pedestrian
  environment: usability and validation.''\ {\em Virtual reality}, 21(3),
  145--152.

\bibitem[\protect\citeauthoryear{}{Sener et~al.\@}{2009}]{sener2009analysis}
Sener, I.~N., Eluru, N., and Bhat, C.~R. (2009).
\newblock ``An analysis of bicycle route choice preferences in texas, us.''\
  {\em Transportation}, 36(5), 511--539.

\bibitem[\protect\citeauthoryear{}{Shaaban
  et~al.\@}{2018}]{shaaban2018analysis}
Shaaban, K., Muley, D., and Mohammed, A. (2018).
\newblock ``Analysis of illegal pedestrian crossing behavior on a major divided
  arterial road.''\ {\em Transportation research part F: traffic psychology and
  behaviour}, 54, 124--137.

\bibitem[\protect\citeauthoryear{}{Shannon}{1948}]{shannon1948mathematical}
Shannon, C.~E. (1948).
\newblock ``A mathematical theory of communication.''\ {\em The Bell system
  technical journal}, 27(3), 379--423.

\bibitem[\protect\citeauthoryear{}{Sharif and
  Oppenheimer}{2021}]{sharif2021effect}
Sharif, M.~A. and Oppenheimer, D.~M. (2021).
\newblock ``The effect of categories on relative encoding biases in
  memory-based judgments.''\ {\em Organizational Behavior and Human Decision
  Processes}, 162, 1--8.

\bibitem[\protect\citeauthoryear{}{Shiferaw
  et~al.\@}{2019}]{shiferaw2019review}
Shiferaw, B., Downey, L., and Crewther, D. (2019).
\newblock ``A review of gaze entropy as a measure of visual scanning
  efficiency.''\ {\em Neuroscience \& Biobehavioral Reviews}, 96, 353--366.

\bibitem[\protect\citeauthoryear{}{Shiferaw
  et~al.\@}{2018}]{shiferaw2018stationary}
Shiferaw, B.~A., Downey, L.~A., Westlake, J., Stevens, B., Rajaratnam, S.~M.,
  Berlowitz, D.~J., Swann, P., and Howard, M.~E. (2018).
\newblock ``Stationary gaze entropy predicts lane departure events in
  sleep-deprived drivers.''\ {\em Scientific reports}, 8(1), 1--10.

\bibitem[\protect\citeauthoryear{}{Shoman and Imine}{2020}]{shoman2020modeling}
Shoman, M. and Imine, H. (2020).
\newblock ``Modeling and simulation of bicycle dynamics.''\ {\em Proc. TRA},
  1--10.

\bibitem[\protect\citeauthoryear{}{Simpson
  et~al.\@}{2003}]{simpson2003investigation}
Simpson, G., Johnston, L., and Richardson, M. (2003).
\newblock ``An investigation of road crossing in a virtual environment.''\ {\em
  Accident Analysis \& Prevention}, 35(5), 787--796.

\bibitem[\protect\citeauthoryear{}{Soares et~al.\@}{2020}]{soares2020driving}
Soares, S., Ferreira, S., and Couto, A. (2020).
\newblock ``Driving simulator experiments to study drowsiness: a systematic
  review.''\ {\em Traffic injury prevention}, 21(1), 29--37.

\bibitem[\protect\citeauthoryear{}{Stelling-Konczak
  et~al.\@}{2018}]{stelling2018study}
Stelling-Konczak, A., Vlakveld, W., Van~Gent, P., Commandeur, J., Van~Wee, B.,
  and Hagenzieker, M. (2018).
\newblock ``A study in real traffic examining glance behaviour of teenage
  cyclists when listening to music: results and ethical considerations.''\ {\em
  Transportation research part F: traffic psychology and behaviour}, 55,
  47--57.

\bibitem[\protect\citeauthoryear{}{Stevens
  et~al.\@}{2013}]{stevens2013preadolescent}
Stevens, E., Plumert, J.~M., Cremer, J.~F., and Kearney, J.~K. (2013).
\newblock ``Preadolescent temperament and risky behavior: Bicycling across
  traffic-filled intersections in a virtual environment.''\ {\em Journal of
  pediatric psychology}, 38(3), 285--295.

\bibitem[\protect\citeauthoryear{}{Stoker
  et~al.\@}{2015}]{stoker2015pedestrian}
Stoker, P., Garfinkel-Castro, A., Khayesi, M., Odero, W., Mwangi, M.~N., Peden,
  M., and Ewing, R. (2015).
\newblock ``Pedestrian safety and the built environment: a review of the risk
  factors.''\ {\em Journal of Planning Literature}, 30(4), 377--392.

\bibitem[\protect\citeauthoryear{}{Stroh}{2017}]{stroh2017design}
Stroh, O. (2017).
\newblock ``The design of an electro-mechanical bicycle for an immersive
  virtual environment.

\bibitem[\protect\citeauthoryear{}{Subramanian
  et~al.\@}{2021}]{subramanian2021pedestrians}
Subramanian, L.~D., O'Neal, E.~E., Roman, A., Sherony, R., Plumert, J.~M., and
  Kearney, J.~K. (2021).
\newblock ``How do pedestrians respond to adaptive headlamp systems in
  vehicles? a road-crossing study in an immersive virtual environment.''\ {\em
  Accident Analysis \& Prevention}, 160, 106298.

\bibitem[\protect\citeauthoryear{}{Sun and Qing}{2018}]{sun2018design}
Sun, C. and Qing, Z. (2018).
\newblock ``Design and construction of a virtual bicycle simulator for
  evaluating sustainable facilities design.''\ {\em Advances in Civil
  Engineering}, 2018.

\bibitem[\protect\citeauthoryear{}{Tapiro et~al.\@}{2016}]{tapiro2016older}
Tapiro, H., Borowsky, A., Oron-Gilad, T., and Parmet, Y. (2016).
\newblock ``Where do older pedestrians glance before deciding to cross a
  simulated two-lane road? a pedestrian simulator paradigm.''\ {\em Proceedings
  of the human factors and ergonomics society annual meeting}, Vol.~60, SAGE
  Publications Sage CA: Los Angeles, CA,  11--15.

\bibitem[\protect\citeauthoryear{}{Tavakoli
  et~al.\@}{2021a}]{tavakoli2021driver}
Tavakoli, A., Kumar, S., Boukhechba, M., and Heydarian, A. (2021a).
\newblock ``Driver state and behavior detection through smart wearables.''\
  {\em arXiv preprint arXiv:2104.13889}.

\bibitem[\protect\citeauthoryear{}{Tavakoli
  et~al.\@}{2021b}]{tavakoli2021harmony}
Tavakoli, A., Kumar, S., Guo, X., Balali, V., Boukhechba, M., and Heydarian, A.
  (2021b).
\newblock ``Harmony: A human-centered multimodal driving study in the wild.''\
  {\em IEEE Access}, 9, 23956--23978.

\bibitem[\protect\citeauthoryear{}{Teixeira et~al.\@}{2020}]{teixeira2020does}
Teixeira, I.~P., da~Silva, A. N.~R., Schwanen, T., Manzato, G.~G.,
  D{\"o}rrzapf, L., Zeile, P., Dekoninck, L., and Botteldooren, D. (2020).
\newblock ``Does cycling infrastructure reduce stress biomarkers in commuting
  cyclists? a comparison of five european cities.''\ {\em Journal of Transport
  Geography}, 88, 102830.

\bibitem[\protect\citeauthoryear{}{Thompson
  et~al.\@}{2013}]{thompson2013impact}
Thompson, L.~L., Rivara, F.~P., Ayyagari, R.~C., and Ebel, B.~E. (2013).
\newblock ``Impact of social and technological distraction on pedestrian
  crossing behaviour: an observational study.''\ {\em Injury prevention},
  19(4), 232--237.

\bibitem[\protect\citeauthoryear{}{Trefzger
  et~al.\@}{2018}]{trefzger2018visual}
Trefzger, M., Blascheck, T., Raschke, M., Hausmann, S., and Schlegel, T.
  (2018).
\newblock ``A visual comparison of gaze behavior from pedestrians and
  cyclists.''\ {\em Proceedings of the 2018 ACM symposium on eye tracking
  research \& applications},  1--5.

\bibitem[\protect\citeauthoryear{}{Tzanavari
  et~al.\@}{2015}]{tzanavari2015effectiveness}
Tzanavari, A., Charalambous-Darden, N., Herakleous, K., and Poullis, C. (2015).
\newblock ``Effectiveness of an immersive virtual environment (cave) for
  teaching pedestrian crossing to children with pdd-nos.''\ {\em 2015 IEEE 15th
  International Conference on Advanced Learning Technologies}, IEEE,  423--427.

\bibitem[\protect\citeauthoryear{}{Tzanavari
  et~al.\@}{2014}]{tzanavari2014user}
Tzanavari, A., Matsentidou, S., Christou, C.~G., and Poullis, C. (2014).
\newblock ``User experience observations on factors that affect performance in
  a road-crossing training application for children using the cave.''\ {\em
  International Conference on Learning and Collaboration Technologies},
  Springer,  91--101.

\bibitem[\protect\citeauthoryear{}{Van~Hoof et~al.\@}{2012}]{van2012comparing}
Van~Hoof, W., Volkaerts, K., O'Sullivan, K., Verschueren, S., and Dankaerts, W.
  (2012).
\newblock ``Comparing lower lumbar kinematics in cyclists with low back pain
  (flexion pattern) versus asymptomatic controls--field study using a wireless
  posture monitoring system.''\ {\em Manual therapy}, 17(4), 312--317.

\bibitem[\protect\citeauthoryear{}{van Paridon et~al.\@}{2019}]{van2019visual}
van Paridon, K.~N., Leivers, H.~K., Robertson, P.~J., and Timmis, M.~A. (2019).
\newblock ``Visual search behaviour in young cyclists: A naturalistic
  experiment.''\ {\em Transportation research part F: traffic psychology and
  behaviour}, 67, 217--229.

\bibitem[\protect\citeauthoryear{}{Van~Veen et~al.\@}{1998}]{van1998navigating}
Van~Veen, H.~A., Distler, H.~K., Braun, S.~J., and B{\"u}lthoff, H.~H. (1998).
\newblock ``Navigating through a virtual city: Using virtual reality technology
  to study human action and perception.''\ {\em Future Generation Computer
  Systems}, 14(3-4), 231--242.

\bibitem[\protect\citeauthoryear{}{Wynne et~al.\@}{2019}]{wynne2019systematic}
Wynne, R.~A., Beanland, V., and Salmon, P.~M. (2019).
\newblock ``Systematic review of driving simulator validation studies.''\ {\em
  Safety science}, 117, 138--151.

\bibitem[\protect\citeauthoryear{}{Xu et~al.\@}{2017}]{xu2017exploring}
Xu, J., Lin, Y., and Schmidt, D. (2017).
\newblock ``Exploring the influence of simulated road environments on cyclist
  behavior.''\ {\em International Journal of Virtual Reality}, 17(3), 15--26.

\bibitem[\protect\citeauthoryear{}{Yang
  et~al.\@}{2019}]{yang2019underreporting}
Yang, H., Cherry, C.~R., Su, F., Ling, Z., Pannell, Z., Li, Y., and Fu, Z.
  (2019).
\newblock ``Underreporting, crash severity and fault assignment of minor
  crashes in china--a study based on self-reported surveys.''\ {\em
  International journal of injury control and safety promotion}, 26(1), 30--36.

\bibitem[\protect\citeauthoryear{}{Zou et~al.\@}{2017}]{zou2017emotional}
Zou, H., Li, N., and Cao, L. (2017).
\newblock ``Emotional response--based approach for assessing the sense of
  presence of subjects in virtual building evacuation studies.''\ {\em Journal
  of Computing in Civil Engineering}, 31(5), 04017028.

\bibitem[\protect\citeauthoryear{}{Zou et~al.\@}{2021}]{zou2021road}
Zou, X., O'Hern, S., Ens, B., Coxon, S., Mater, P., Chow, R., Neylan, M., and
  Vu, H.~L. (2021).
\newblock ``On-road virtual reality autonomous vehicle (vrav) simulator: An
  empirical study on user experience.''\ {\em Transportation Research Part C:
  Emerging Technologies}, 126, 103090.

\end{thebibliography}

\end{document}